\documentclass[journal]{IEEEtran}

\usepackage{cite} 
\usepackage{CJKutf8}
\usepackage{algorithm}
\usepackage{algorithmic}
\usepackage{amsmath}
\DeclareMathOperator{\diag}{diag}
\DeclareMathOperator*{\argmin}{arg\,min}
\usepackage{graphicx}
\usepackage{amssymb}
\usepackage{stfloats}
\usepackage{bm}
\usepackage{booktabs}
\usepackage{array}
\usepackage{tabularx}
\usepackage{caption}
\usepackage{epsfig,upgreek}
\usepackage[caption=false,font=footnotesize]{subfig}
\usepackage{hyperref}
\usepackage{xcolor}  
\hypersetup{
	colorlinks=true,
	linkcolor=black,         
	citecolor=black,         
	urlcolor=black          
}

\ifCLASSINFOpdf

\else

\fi

\begin{document}
\title{Widely Linear-Enhanced IQ Imbalance Interference Cancellation for AFDM-FTN Systems}

\author{
\IEEEauthorblockN{Xianle Dai,  Bowei Xue, Qu Luo,~\IEEEmembership{Member,~IEEE,} Jianguo Li,~\IEEEmembership{Member,~IEEE,} Neng Ye,~\IEEEmembership{Senior Member,~IEEE,} Kai Yang,~\IEEEmembership{Member,~IEEE,}  and Pei Xiao,~\IEEEmembership{Senior Member,~IEEE}}
\thanks{Xianle Dai and Kai Yang are with the School of Information and Electronics, Beijing Institute of Technology, Beijing 100081, China (e-mail: daixl212@bit.edu.cn, yangkai@ieee.org).}% <-this % stops a space
\thanks{Qu Luo and Pei Xiao are with the 5G/6G Innovation Centre, University of Surrey, GU2 7XH Guildford, U.K. (e-mail: \{q.u.luo, p.xiao\}@surrey.ac.uk).}
\thanks{Bowei Xue, Jianguo Li and Neng Ye are with the School of Cyberspace Science and Technology, Beijing Institute of Technology, Beijing 100081, China (e-mail: \{boweixue, jianguoli, ianye\}@bit.edu.cn).}
}

% \author{First A. Author, \IEEEmembership{Fellow, IEEE}, Second B. Author, and Third C. Author, Jr., \IEEEmembership{Member, IEEE}
% \thanks{This paragraph of the first footnote will contain the date on 
% which you submitted your brief for review. It will also contain support 
% information, including sponsor and financial support acknowledgment. For 
% example, ``This work was supported in part by the U.S. Department of 
% Commerce under Grant BS123456.'' }
% \thanks{The next few paragraphs should contain 
% the authors' current affiliations, including current address and e-mail. For 
% example, F. A. Author is with the National Institute of Standards and 
% Technology, Boulder, CO 80305 USA (e-mail: author@boulder.nist.gov). }
% \thanks{S. B. Author, Jr., was with Rice University, Houston, TX 77005 USA. He is 
% now with the Department of Physics, Colorado State University, Fort Collins, 
% CO 80523 USA (e-mail: author@lamar.colostate.edu).}
% \thanks{T. C. Author is with 
% the Electrical Engineering Department, University of Colorado, Boulder, CO 
% 80309 USA, on leave from the National Research Institute for Metals, 
% Tsukuba, Japan (e-mail: author@nrim.go.jp).}}

\maketitle
%%===========================================Abstract================================================%%
\begin{abstract}
Internet of Things (IoT) networks require reliable and spectrally efficient wireless links to support massive device connectivity and high mobility applications. To meet these requirements, this paper investigates a widely linear (WL)-empowered receiver design for affine frequency division multiplexing (AFDM) with faster-than-Nyquist (FTN) signaling under practical receiver in phase and quadrature imbalance (IQI). The considered AFDM-FTN system couples the direct and image branches with FTN induced intersymbol interference (FTN-ISI) and delay-Doppler spreading in the affine domain.  
To address this coupled interference structure, a unified conjugate augmented input output model is established for IQI impaired AFDM-FTN systems. Based on this model, a  WL minimum mean square error receiver with parallel interference cancellation (WL-MMSE-PIC) is developed,    
where WL-MMSE preprocessing is followed by the proposed hard decision PIC (HD-PIC) and damped soft decision PIC (SD-PIC). The proposed receiver is further analyzed in terms of the local stability of HD-PIC, a genie aided performance reference under ideal residual interference cancellation, and the impact of the AFDM chirp parameter  on the post equalization interference distribution. 
Simulation results show that the proposed WL-MMSE-HD-PIC and WL-MMSE-SD-PIC receivers achieve significant error rate performance gains over the  conventional linear-MMSE and WL-MMSE baselines. 
The results further show that FTN signaling provides a nominal spectral-efficiency gain, while residual-coupling-based chirp-parameter selection reduces the post-equalization residual coupling and improves detection reliability for high-mobility IoT links.
\end{abstract}

%%===========================================Keywords================================================%%
\begin{IEEEkeywords}
AFDM, FTN signaling, IQI, WL-MMSE, PIC, doubly selective channels.
\end{IEEEkeywords}

%%=========================================Introduction===============================================%%
\section{Introduction}
\label{sec:introduction}

Internet of Things (IoT) networks impose stringent requirements on link reliability and spectrum utilization. In this context, next generation wireless networks are expected to support massive and heterogeneous IoT connectivity in high mobility scenarios such as high speed railways, unmanned aerial vehicles, and satellite communications~\cite{AFDMSCMA,nguyen2022iot6g,AFDMstd}. In these scenarios, severe  Doppler spreading destroys the orthogonality of legacy orthogonal frequency division multiplexing (OFDM), leading to significant performance degradation~\cite{zhang2026unified_multicarrier}. This limitation motivates the development of Doppler resilient waveforms. Orthogonal time frequency space (OTFS) modulates information symbols in the delay-Doppler domain and has been investigated as a candidate waveform for high mobility channels~\cite{aldababsa2024otfs_survey,xiao2022iot}. Orthogonal chirp division multiplexing (OCDM), building  upon the discrete Fresnel transform (DFnT), modulates the information symbols using a class of orthogonal chirp subcarriers~\cite{zhang2026unified_multicarrier}. However, OCDM may not fully exploit the channel characteristics when its fixed chirp rate is mismatched with the specific delay-Doppler profile. More recently, affine frequency division multiplexing (AFDM) has been proposed for doubly selective channels in the discrete affine Fourier transform (DAFT) domain, where its adjustable chirp parameters provide additional flexibility in shaping the delay-Doppler dispersion~\cite{yin2025afdm_overview,luoafdmisac}. By properly selecting the chirp parameters, AFDM can  transform a doubly selective channel into a structured, quasistatic equivalent channel in the DAFT domain, thereby improving the exploitation of delay-Doppler diversity. 
Receiver processing has also been studied to improve channel acquisition and parameter estimation~\cite{chen2025structured_tensor,chen2026tensor_isac}.
These studies indicate that chirp domain and delay Doppler domain waveforms can effectively improve transmission reliability in high mobility environments.

To enhance spectral efficiency in high mobility scenarios, AFDM has been integrated with nonorthogonal multiple access~\cite{AFDMSCMA}, index modulation~\cite{tao2025afdm_im,liu2026mm_afdm_im,ZhuIMAFDM}, and nonorthogonal waveform design~\cite{yi2026nonorth_afdm}. For example, the authors in~\cite{AFDMSCMA} proposed AFDM-assisted sparse code multiple access, referred to as AFDM-SCMA, to support massive access in high mobility scenarios. In addition, AFDM with index modulation conveys information through both constellation symbols and chirp subcarrier activation patterns~\cite{tao2025afdm_im,liu2026mm_afdm_im}, whereas nonorthogonal AFDM further enhances spectral efficiency through controlled subcarrier overlap~\cite{yi2026nonorth_afdm}. Recent work has also developed a low-complexity vector-by-vector detector for AFDM index modulation by exploiting the sparse channel structure~\cite{wang2025vvafdm}. In parallel, faster-than-Nyquist (FTN) signaling offers a complementary approach by transmitting symbols at intervals shorter than the Nyquist interval~\cite{mazo1975ftn}. Recent studies have revisited FTN signaling from the perspectives of waveform design~\cite{jaffal2022ftn_pulse}, pulse shaping~\cite{li2026ftn_overview}, and receiver processing~\cite{tong2025ftn_iqi_ddsic}. 
Since multicarrier modulation enables domainwise signal processing and FTN signaling increases the nominal symbol rate under a fixed bandwidth~\cite{li2020ftn_equalization,zhang2026unified_multicarrier}, integrating FTN with AFDM provides a possible approach to jointly address high mobility and spectral efficiency requirements.
However, FTN time compression introduces controlled intersymbol interference (ISI), which is further coupled with AFDM domain delay-Doppler spreading. The resulting coupling  makes receiver design more challenging.

Despite these advantages, the practical implementation of AFDM-FTN systems is inevitably affected by hardware impairments. In high frequency and energy constrained transceivers, finite resolution analog-to-digital converters (ADCs)~\cite{meng2026afdm_adc}, oscillator phase noise and carrier frequency offset (CFO)~\cite{bemani2023afdm_twc,yin2025afdm_overview}, nonlinear power amplification~\cite{chen2026clipped_otfs}, and I/Q imbalance (IQI)~\cite{sui2026mimo_afdm_hwi} can distort the received waveform. Among these impairments, receiver IQI is particularly critical in direct conversion architectures, where amplitude and phase mismatches between the in-phase and quadrature branches generate image leakage coupled with the desired signal~\cite{tarighat2005iqi,tubbax2005iqi}. In AFDM-FTN transmission, this image component is superimposed on the interference induced by FTN time compression and doubly selective propagation, thereby giving rise to a coupled detection problem.

Extensive studies have investigated receiver design and performance analysis for AFDM and FTN systems. For AFDM reception, the authors in~\cite{luo2026jsg_afdm} developed a joint sparse-graph detector that exploits both the structured sparsity of the AFDM channel and the LDPC decoding graph to enable joint detection and decoding in MIMO-AFDM systems. 
In addition, an MMSE based chirp parameter selection criterion was proposed in~\cite{li2024chirp_mmse_afdm} to reshape the post equalization interference distribution and improve detection performance. Receiver IQI compensation in AFDM has also been investigated through a cascade strategy for joint transmitter and receiver IQI compensation~\cite{liu2026afdm_iqi}. 
For FTN signaling, constrained-capacity analysis has characterized the fundamental limits associated with time domain compression~\cite{rusek2009capacity}, while reduced complexity sequence detectors have been developed to mitigate FTN induced ISI~\cite{liveris2003ftn,prlja2012ftn_receiver}. Frequency-domain equalization, variational inference-based detection, and probabilistic data association have further enabled efficient FTN detection with hard and soft outputs~\cite{sugiura2015ftn_fde,yuan2016ftn_doubly,li2020ftn_equalization}. Moreover, to mitigate IQI induced degradation in FTN systems, a joint ISI-IQI cancellation scheme was proposed in~\cite{tong2025ftn_iqi_ddsic} to account for the interaction between FTN induced ISI and IQI induced image interference.

Nevertheless, existing receivers are primarily designed for either AFDM or FTN specific interference structures~\cite{tong2025ftn_iqi_ddsic,wang2025vvafdm}. The proposed AFDM-FTN architecture introduces a distinct coupling between DAFT domain channel dispersion and FTN induced ISI, which is further complicated by practical receiver IQI. Consequently, conventional AFDM or FTN detectors may not be directly applicable to the considered system, thereby necessitating a unified receiver capable of jointly mitigating these intertwined interference components.

Motivated by this challenge, this paper develops a receiver framework for spectrally efficient AFDM-FTN transmission over high mobility channels in the presence of  IQI. Specifically, the desired signal, IQI induced image component, FTN induced ISI, and delay–Doppler channel coupling are incorporated into a unified conjugate augmented input–output model. Based on this model, an augmented widely linear minimum mean square error (WL-MMSE) equalizer is developed to jointly suppress IQI induced image interference and AFDM-FTN coupling. Since residual off diagonal interference remains after linear equalization, hard decision PIC (HD-PIC) and soft decision PIC (SD-PIC) schemes are further designed to iteratively reconstruct and cancel the residual interference.

The main contributions of this paper are summarized as follows.
\begin{itemize}
\item We propose an AFDM-FTN transmission system that combines the high mobility robustness of AFDM with the spectral efficiency enhancement offered by FTN signaling. The practical impact of receiver IQI is also taken into account. Accordingly, a unified conjugate augmented input output model is established by incorporating CPP insertion, FTN pulse shaping, doubly selective propagation, matched filtering, AFDM demodulation, and receiver IQI, thereby characterizing the coupled effects of FTN induced ISI, delay-Doppler spreading, and IQI induced image interference.

\item We propose a low-complexity WL-MMSE-PIC receiver to address IQI induced conjugate distortion and the coupled interference arising from FTN signaling and AFDM transmission. The WL-MMSE preprocessing jointly exploits the direct and image branches, and the resulting post equalization matrix is decomposed into diagonal main gain and off diagonal residual coupling components. Based on this decomposition, HD-PIC and SD-PIC are developed to iteratively reconstruct and cancel the residual post equalization interference using different feedback strategies. Specifically, HD-PIC employs hard constellation decisions, whereas SD-PIC uses posterior means together with variance and pseudo variance information in a bivariate likelihood, with damping applied to mitigate unreliable feedback.

\item We provide a rigorous theoretical analysis of the proposed WL-MMSE-PIC receiver, covering local stability, a genie aided performance reference, and AFDM-FTN chirp parameter selection. The HD-PIC stability condition is related to the normalized residual coupling matrix, while SD-PIC is interpreted through posterior variance, pseudo variance, and damping control. The AFDM chirp parameter $c_1$ is further connected to the post equalization interference distribution through a residual leakage based selection criterion.

\item We show that the proposed receivers achieve favorable BER performance and iterative convergence under different chirp grid selections and FTN compression factors, while maintaining a balanced complexity performance tradeoff. The results show that HD-PIC achieves effective residual cancellation with relatively low per iteration complexity, whereas SD-PIC mitigates feedback induced error propagation through posterior mean feedback, covariance and pseudo covariance information, and damping.
\end{itemize}

The remainder of this paper is organized as follows. Section~\ref{sec:system_model} presents the AFDM-FTN system model under doubly selective channel and receiver IQI. Section~\ref{sec:ReceiverDesign} develops the proposed WL-MMSE-PIC receiver and analyzes its computational complexity. Section~\ref{sec:4} provides the local stability analysis, genie aided analytical reference, and parameter selection discussion. Section~\ref{sec:simulation} presents simulation results and discussions. Finally, Section~\ref{sec:conclusion} concludes this paper.

\textit{Notation:} Bold uppercase, bold lowercase, and lowercase italic letters denote matrices, vectors, and scalars, respectively. The notation $\mathbf X\in\mathbb C^{m\times n}$ denotes an $m\times n$ complex valued matrix. The superscripts $(\cdot)^T$, $(\cdot)^H$, and $(\cdot)^*$ denote transpose, Hermitian transpose, and complex conjugation. The operators $\mathbb E[\cdot]$, $\diag(\cdot)$, $\lfloor\cdot\rfloor$, and $\argmin$ denote expectation, diagonal operation, floor function, and minimization argument, respectively. The symbols $\mathcal{CN}(\cdot,\cdot)$, $\mathcal O(\cdot)$, and $\otimes$ denote the complex Gaussian distribution, asymptotic complexity order, and Kronecker product, respectively. The notations $[\mathbf x]_k$ and $[\mathbf X]_{m,n}$ denote the $k$th entry of vector $\mathbf x$ and the $(m,n)$th entry of matrix $\mathbf X$, respectively. The notations $\|\cdot\|_2$ and $\|\cdot\|_\infty$ denote the Euclidean norm and the infinity norm, where $\|\cdot\|_\infty$ is the maximum absolute entry for a vector and the maximum absolute row sum for a matrix. For a scalar $a$, the notation $[a]_N$ denotes $a$ modulo $N$. The symbols $\mathbf 0$ and $\mathbf I_N$ denote an all zero vector or matrix with compatible dimensions and the identity matrix of size indicated by the subscript, respectively.

The operator $\operatorname{vec}(\cdot)$ denotes columnwise vectorization.

%%=======================================Chapter 2 System Model===============================================%%
\section{System Model}
\label{sec:system_model}

\begin{figure*}[!t] 
    \centerline{\includegraphics[width=0.9\textwidth]{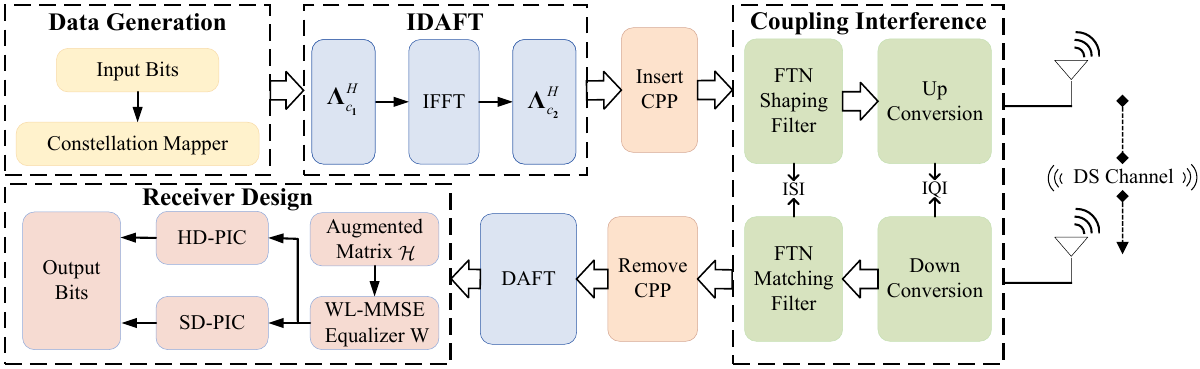}}
	\caption{Transceiver block diagram of the considered AFDM-FTN system with receiver IQI.}
	\label{fig:system_model}
\end{figure*}

This section presents the system model of the proposed AFDM-FTN system in the presence of receiver IQI, as illustrated in Fig.~\ref{fig:system_model}.

%===========================================Figure system_model=========================================%

%=======================================================================================================%

\subsection{Transmit Signal Model}
Let $\mathcal S$ denote the $M$-ary quadrature amplitude modulation ($M$-QAM) alphabet, and $N$ denote the number of subcarriers. The data vector $\mathbf d=[d_0,\ldots,d_{N-1}]^T\in\mathcal S^N$ is assumed to be circularly symmetric and satisfies $\mathbb{E}[\mathbf d\mathbf d^T]=\mathbf 0$ and $\mathbb{E}[\mathbf d\mathbf d^H]=\sigma_d^2\mathbf I_N$, where $\sigma_d^2$ denotes the average symbol energy. After AFDM modulation, CPP insertion, and FTN pulse shaping, the transmitted vector is expressed as
\begin{equation}
 \scalebox{1}{$%
		\begin{aligned}
			\mathbf x=\mathbf T\mathbf d,
		\end{aligned}
		$}
	\label{eq:tx_signal}
\end{equation}
where $\mathbf T\triangleq\mathbf G\mathbf C_{\rm cpp}\mathbf A^H$ is the overall AFDM-FTN transmit matrix of size $N_s\times N$.
The DAFT matrix is $\mathbf A=\boldsymbol\Lambda_{c_2}\mathbf F\boldsymbol\Lambda_{c_1}\in\mathbb C^{N\times N}$, where $c_1$ and $c_2$ are the AFDM chirp parameters~\cite{bemani2023afdm_twc}. The diagonal chirp matrix is $\boldsymbol\Lambda_c=\diag(e^{-j2\pi c n^2})$, $n=0,1,\ldots,N-1$, and $\mathbf F$ is the discrete Fourier transform (DFT) matrix with entry $e^{-j2\pi mn/N}/\sqrt{N}$ at the $(m,n)$th position.

Let $L_{\rm cpp}$ denote the CPP length and $N_s=N+L_{\rm cpp}$. The CPP insertion matrix is
\begin{equation}
	\scalebox{1}{$%
		\begin{aligned}
\mathbf C_{\rm cpp}=
\begin{bmatrix}
\boldsymbol\Phi_{\rm cpp}\mathbf J_t\\
\mathbf I_N
\end{bmatrix}
\in\mathbb{C}^{N_s\times N},
		\end{aligned}
		$}
	\label{eq:cpp_insertion}
\end{equation}
where $\mathbf J_t=[\mathbf 0_{L_{\rm cpp}\times (N-L_{\rm cpp})} \, \mathbf I_{L_{\rm cpp}}]$ extracts the last $L_{\rm cpp}$ samples, and $\boldsymbol\Phi_{\rm cpp}=\diag(e^{-j2\pi c_1(N^2+2N(\ell-L_{\rm cpp}))})$, $\ell=0,\ldots,L_{\rm cpp}-1$, is the CPP phase rotation matrix. 
The CPP length is chosen to be no smaller than the maximum integer path delay.

The matrix $\mathbf G\in\mathbb C^{N_s\times N_s}$ denotes the FTN pulse shaping matrix and represents the controlled ISI intentionally introduced by time domain compression~\cite{anderson2013ftn,jaffal2022ftn_pulse}.
For a shaping pulse $g(t)$ sampled at the compressed symbol interval $\alpha T$, the blockwise circular FTN filtering matrix is modeled by
\begin{equation}
\scalebox{1}{$%
		\begin{aligned}
			{\relax[\mathbf G]_{m,n}=g \left(\left[(m-n)\bmod N_s\right]\alpha T\right),}\\
			{m,n=0,1,\ldots,N_s-1,}
		\end{aligned}
		$}
	\label{eq:ftn_matrix}
\end{equation}
where $T$ is the Nyquist symbol interval. The transmit pulse is truncated to $|t|\leq K_g\alpha T$, where $K_g$ is the pulse half span. The truncated pulse is periodically extended over the block duration $N_s\alpha T$ to construct the circular convolution matrix $\mathbf G$.
In parameter selection, the compression factor is chosen from $\alpha\in\mathcal A$, where $\mathcal A\subset(0,1]$ denotes the candidate set of FTN compression factors~\cite{anderson2013ftn}.

\subsection{Channel and I/Q Imbalance}
The propagation channel is modeled as a doubly selective channel, whose matrix representation is as follows
\begin{equation}
	\scalebox{1}{$%
		\begin{aligned}
\mathbf H_{\rm ch}=\sum_{p=1}^{P}h_p\boldsymbol\Delta(\nu_p)\boldsymbol\Pi(\tau_p),
		\end{aligned}
		$}
	\label{eq:ds_channel}
\end{equation}
where $P$ is the number of propagation paths, and $h_p$, $\nu_p$, and $\tau_p$ denote the complex fading coefficient, normalized Doppler index, and integer delay tap of the $p$th path, respectively. 
% The normalized Doppler index is related to the physical Doppler frequency by $\nu_p=f_{D,p}N_sT_s$, where $f_{D,p}$ is the physical Doppler frequency and $T_s$ is the sampling interval. 
In \eqref{eq:ds_channel}, $\boldsymbol\Delta(\nu_p)$ models the Doppler with its expression given by 
\begin{equation}
	\scalebox{1}{$%
		\begin{aligned}
\boldsymbol\Delta(\nu_p)=\diag\left(e^{-j2\pi\nu_p n/N_s}\right),\quad n=0,1,\ldots,N_s-1.
		\end{aligned}
		$}
	\label{eq:doppler_matrix}
\end{equation}
Note that $\nu_p=f_{D,p}N_sT_s$ is the normalized  Doppler frequency, where $f_{D,p}$ is the  analog Doppler frequency  and $T_s$ is the sampling interval. 
$\boldsymbol\Pi(\tau_p)$ is the circular delay shift matrix whose $(n,m)$th element is defined as
\begin{equation}
	\scalebox{1}{$%
		\begin{aligned}
\relax[\boldsymbol\Pi(\tau_p)]_{n,m}=\begin{cases}
1, & n=(m+\tau_p)\bmod N_s,\\
0, & \text{otherwise}.
\end{cases}
		\end{aligned}
		$}
	\label{eq:delay_matrix}
\end{equation}
After propagation through the doubly selective channel, the received signal before RF front end imbalance is
\begin{equation}
	\scalebox{1}{$%
		\begin{aligned}
\mathbf r=\mathbf H_{\rm ch}\mathbf x+\mathbf w,
		\end{aligned}
		$}
	\label{eq:rx_before_iqi}
\end{equation}
where $\mathbf w\sim\mathcal{CN}(\mathbf 0,\sigma_w^2\mathbf I_{N_s})$ is proper complex Gaussian noise, and $\sigma_w^2$ denotes the noise variance.

Practical direct conversion receivers may suffer from receiver I/Q imbalance during down conversion, which introduces image leakage coupled to the desired signal.
Let $\mathbf U$ and $\mathbf V$ denote the equivalent direct response matrix and image response matrix induced by receiver IQI, respectively. After down conversion (DC), the received signal is modeled by
\begin{equation}
	\scalebox{1}{$%
		\begin{aligned}
			\tilde{\mathbf r}=\mathbf U\mathbf r+\mathbf V\mathbf r^*.
		\end{aligned}
		$}
	\label{eq:iqi_model}
\end{equation}
For receiver IQI that is frequency independent, the IQI response matrices are scalar multiples of the identity, i.e., $\mathbf U=\mu\mathbf I_{N_s}$ and $\mathbf V=\nu\mathbf I_{N_s}$. The coefficients $\mu$ and $\nu$ follow the standard receiver IQI parameterization based on amplitude and phase mismatches~\cite{tarighat2005iqi, tubbax2005iqi}, with
\begin{subequations}
\label{eq:iqi_coeff}
\begin{align}
\mu=\cos\left(\frac{\phi}{2}\right)+j\epsilon\sin\left(\frac{\phi}{2}\right),\label{eq:iqi_coeff_mu}\\
\nu=\epsilon\cos\left(\frac{\phi}{2}\right)-j\sin\left(\frac{\phi}{2}\right),\label{eq:iqi_coeff_nu}
\end{align}
\end{subequations}
where $\epsilon$ and $\phi$ denote the amplitude and phase mismatches, respectively. The ideal I/Q matching case gives $\mu=1$ and $\nu=0$. In addition, the image rejection ratio (IRR), which is commonly used in RF front end characterization and IQI impaired receiver simulations, is defined to quantify the power ratio between the real and image components as
\begin{equation}
	\scalebox{1}{$%
		\begin{aligned}
{\rm IRR}=10\log_{10}\frac{|\mu|^2}{|\nu|^2}.
		\end{aligned}
		$}
	\label{eq:irr}
\end{equation}
Combining \eqref{eq:rx_before_iqi} and \eqref{eq:iqi_model} gives the IQI impaired signal
\begin{equation}
	\scalebox{1}{$%
		\begin{aligned}
\tilde{\mathbf r}=\mathbf U\mathbf H_{\rm ch}\mathbf T\mathbf d+\mathbf V(\mathbf H_{\rm ch}\mathbf T)^*\mathbf d^*+\tilde{\mathbf w},
		\end{aligned}
		$}
	\label{eq:iqi_rx_expanded}
\end{equation}
where $\tilde{\mathbf w}=\mathbf U\mathbf w+\mathbf V\mathbf w^*$ is the equivalent IQI impaired noise.

\subsection{Equivalent Received Signal Model}
After FTN matched filtering, CPP removal, and AFDM demodulation, the receiver observation is $\mathbf y_{\rm eq}=\mathbf B_{\rm rx}\tilde{\mathbf r}$, where $\mathbf B_{\rm rx}\triangleq\mathbf A\mathbf R_{\rm cpp}\mathbf M$ is the receiver side processing matrix, $\mathbf M=\mathbf G^H$ denotes the matched filtering matrix, and $\mathbf R_{\rm cpp}=[\mathbf 0_{N\times L_{\rm cpp}}\,\mathbf I_N]$ removes the CPP samples.
The receive processing matrix $\mathbf B_{\rm rx}\in\mathbb C^{N\times N_s}$ applies matched filtering, CPP removal, and DAFT demodulation in that order.

According to \eqref{eq:iqi_rx_expanded}, the receiver processed observation can be represented  as
\begin{equation}
	\scalebox{1}{$%
		\begin{aligned}
\mathbf y_{\rm eq}=\mathbf H_{\rm eq}\mathbf d+\mathbf G_{\rm eq}\mathbf d^*+\mathbf n_{\rm eq},
		\end{aligned}
		$}
	\label{eq:augmented}
\end{equation}
where $\mathbf H_{\rm eq}$ and $\mathbf G_{\rm eq}$ denote the real and image equivalent channels, respectively, and the corresponding channel and noise terms are
\begin{subequations}
\label{eq:Heq}
\begin{align}
\mathbf H_{\rm eq}
&=
\mathbf B_{\rm rx}\mathbf U\mathbf H_{\rm ch}\mathbf T,
\label{eq:Geq}\\
\mathbf G_{\rm eq}
&=
\mathbf B_{\rm rx}\mathbf V
(\mathbf H_{\rm ch}\mathbf T)^*,
\\
\mathbf n_{\rm eq}
&=
\mathbf B_{\rm rx}
\left(
\mathbf U\mathbf w+\mathbf V\mathbf w^*
\right).
\label{eq:neq}
\end{align}
\end{subequations}
Equation~\eqref{eq:augmented} contains the direct data component, the IQI induced image component, and the equivalent noise term.
The equivalent channels in \eqref{eq:Heq} incorporate transmit pulse shaping through $\mathbf T$ and receive matched filtering through $\mathbf B_{\rm rx}$, with their joint effect determining the FTN-ISI in the processed observation.
Since receiver IQI destroys the I/Q symmetry and yields an improper conjugate coupled observation, widely linear processing is adopted in Section~\ref{sec:ReceiverDesign} to exploit both the received signal and its conjugate for joint direct and image component suppression.

%%=======================================Chapter 3 Receiver Design============================================%%
\section{WL-MMSE-PIC Receiver Design}
\label{sec:ReceiverDesign}

In this section, we present the proposed WL-MMSE-PIC detector for the considered AFDM-FTN system with receiver IQI. We start by organizing the real and image components into a conjugate augmented model for WL-MMSE preprocessing. Subsequently, the post equalization relation is separated into a main gain component and a residual coupling component, which provides the basis for the HD-PIC and SD-PIC designs. Furthermore, the computational complexity of the considered detection methods is analyzed.

\subsection{WL-MMSE Preprocessing}
Since \eqref{eq:augmented} contains both $\mathbf d$ and $\mathbf d^*$, WL-MMSE preprocessing is formulated in a conjugate augmented domain. The augmented observation, transmitted symbol, and noise vectors are defined by \cite{WLmimo,WLmimo2}
\begin{equation}
	\scalebox{1}{$%
		\begin{aligned}
\mathbf y&\triangleq
\begin{bmatrix}
\mathbf y_{\rm eq}\\
\mathbf y_{\rm eq}^*
\end{bmatrix},\quad
\mathbf s\triangleq
\begin{bmatrix}
\mathbf d\\
\mathbf d^*
\end{bmatrix},\quad
\mathbf n\triangleq
\begin{bmatrix}
\mathbf n_{\rm eq}\\
\mathbf n_{\rm eq}^*
\end{bmatrix}.
		\end{aligned}
		$}
	\label{eq:aug_vectors}
\end{equation}
The processed noise statistics $\mathbf R_{n_{\rm eq}},\mathbf C_{n_{\rm eq}}\in\mathbb C^{N\times N}$ are defined in the processed observation domain, and $\mathbf R_n\in\mathbb C^{2N\times2N}$ is the covariance in the conjugate augmented domain.
The corresponding augmented noise covariance takes the form
\begin{equation}
	\scalebox{1}{$%
		\begin{aligned}
\mathbf R_n
=\mathbb E[\mathbf n\mathbf n^H]
=
\begin{bmatrix}
\mathbf R_{n_{\rm eq}} & \mathbf C_{n_{\rm eq}}\\
\mathbf C_{n_{\rm eq}}^* & \mathbf R_{n_{\rm eq}}^*
\end{bmatrix},
		\end{aligned}
		$}
	\label{eq:aug_noise_cov}
\end{equation}
The effective noise after receiver IQI is given by $\tilde{\mathbf w}=\mathbf U\mathbf w+\mathbf V\mathbf w^*$. Since $\mathbf w$ is proper white noise, $\mathbb E[\mathbf w\mathbf w^H]=\sigma_w^2\mathbf I_{N_s}$ and $\mathbb E[\mathbf w\mathbf w^T]=\mathbf 0$. Its covariance and pseudo covariance are given by
{
\begin{subequations}
\begin{equation}
\mathbf R_{\tilde w}=\mathbb E[\tilde{\mathbf w}\tilde{\mathbf w}^H]
=\sigma_w^2(\mathbf U\mathbf U^H+\mathbf V\mathbf V^H).
\end{equation}
\begin{equation}
\mathbf C_{\tilde w}=\mathbb E[\tilde{\mathbf w}\tilde{\mathbf w}^T]
=\sigma_w^2(\mathbf U\mathbf V^T+\mathbf V\mathbf U^T).
\end{equation}
\end{subequations}
}
Applying $\mathbf B_{\rm rx}$ transforms these moments into the processed observation domain. Specifically, the processed covariance and pseudo covariance are respectively given by
{
\begin{subequations}
\begin{equation}
\mathbf R_{n_{\rm eq}}=\mathbb E[\mathbf n_{\rm eq}\mathbf n_{\rm eq}^H]
=\mathbf B_{\rm rx}\mathbf R_{\tilde w}\mathbf B_{\rm rx}^H.
\label{eq:eq_noise_cov}
\end{equation}
\begin{equation}
\mathbf C_{n_{\rm eq}}=\mathbb E[\mathbf n_{\rm eq}\mathbf n_{\rm eq}^T]
=\mathbf B_{\rm rx}\mathbf C_{\tilde w}\mathbf B_{\rm rx}^T.
\label{eq:eq_noise_pcov}
\end{equation}
\end{subequations}
}
The pseudo covariance $\mathbf C_{n_{\rm eq}}$ is generally nonzero under receiver IQI, and the augmented noise covariance in \eqref{eq:aug_noise_cov} is retained in the WL-MMSE design.
With these definitions, \eqref{eq:augmented} can be expressed in an augmented model as
\begin{equation}
	\scalebox{1}{$%
		\begin{aligned}
			\mathbf y=\mathcal H\mathbf s+\mathbf n,
		\end{aligned}
		$}
	\label{eq:aug_model}
\end{equation}
where $\mathcal H$ is the augmented equivalent channel, i.e., 
\begin{equation}
	\scalebox{1}{$%
		\begin{aligned}
\mathcal H=\begin{bmatrix}
\mathbf H_{\rm eq} & \mathbf G_{\rm eq}\\
\mathbf G_{\rm eq}^* & \mathbf H_{\rm eq}^*
\end{bmatrix}.
		\end{aligned}
		$}
	\label{eq:aug_channel}
\end{equation}
% Since each candidate $\mathbf d$ uniquely determines the augmented vector $\mathbf s$, the optimum joint detection criterion associated with \eqref{eq:aug_model} is
% \begin{equation}
% 	\scalebox{1}{$%
% 		\begin{aligned}
% \hat{\mathbf d}_{\rm ML}
% &=\argmin_{\mathbf d\in\mathcal S^N}
% \left(
% \mathbf y-\mathcal H\mathbf s
% \right)^H
% \mathbf R_n^{-1}
% \left(
% \mathbf y-\mathcal H\mathbf s
% \right).
% 		\end{aligned}
% 		$}
% 	\label{eq:problem}
% \end{equation}
% The exhaustive search complexity of \eqref{eq:problem} scales as $\mathcal O(|\mathcal S|^N)$, which is prohibitive for practical AFDM-FTN systems~\cite{prlja2012ftn_receiver,kulhandjian2020ftn_pda}. Therefore, a lower complexity receiver is required to suppress the coupled interference embedded in $\mathcal H$.

Following the widely linear estimation framework in~\cite{picinbono1995widely,mattera2008wl_iqi}, let $\mathbf W\in\mathbb{C}^{2N\times 2N}$ denote the WL-MMSE equalizer matrix that linearly estimates $\mathbf s$ from $\mathbf y$. With the estimation error defined as  $\mathbf e=\mathbf s-\mathbf W\mathbf y$, the MMSE orthogonality condition is then expressed as $\mathbb{E}\left[\mathbf e\mathbf y^H\right]=\mathbf 0$.
The corresponding WL-MMSE solution is given by
\begin{equation}
	\scalebox{1}{$%
		\begin{aligned}
			\mathbf W
			=\mathbf R_s\mathcal H^H
			\left(\mathcal H\mathbf R_s\mathcal H^H+\mathbf R_n\right)^{-1},
		\end{aligned}
		$}
	\label{eq:wlmmse}
\end{equation}
where $\mathbf R_s=\mathbb{E}[\mathbf s\mathbf s^H]$ and $\mathbf R_n$ are the augmented symbol and noise covariance matrices, respectively.
The WL-MMSE filter in \eqref{eq:wlmmse} uses the augmented noise covariance $\mathbf R_n$, whose diagonal and off diagonal blocks contain the processed covariance and pseudo covariance, respectively.
Under the assumed zero mean, independent, and proper modulation symbols, $\mathbf R_s=\sigma_d^2\mathbf I_{2N}$.
% \begin{equation}
% 	\scalebox{1}{$%
% 		\begin{aligned}
% \bar{\mathbf R}_n=\sigma_n^2\mathbf I_{2N}.
% 		\end{aligned}
% 		$}
% 	\label{eq:white_noise_approx}
% \end{equation}
% The WL-MMSE output used for PIC initialization is
% \begin{equation}
% 	\scalebox{1}{$%
% 		\begin{aligned}
% \mathbf z^{(0)}=\mathbf W\mathbf y.
% 		\end{aligned}
% 		$}
% 	\label{eq:wlmmse_output}
% \end{equation}

\begin{figure*}[!t] 
    \centerline{\includegraphics[width=0.9\textwidth]{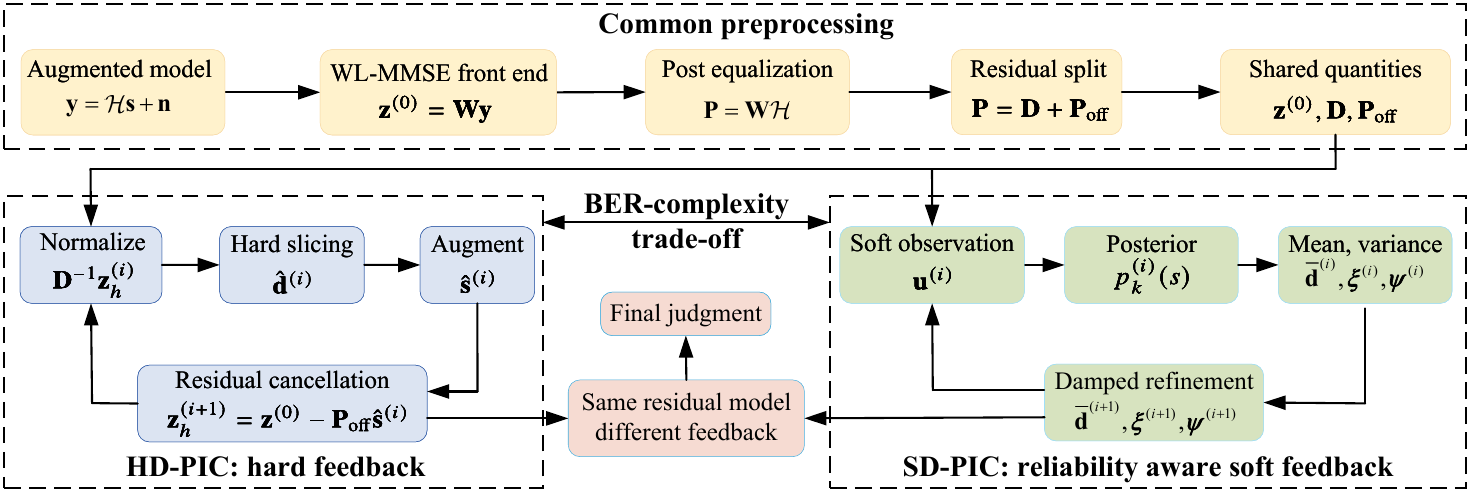}}
		\caption{Flowchart of the proposed WL-MMSE-PIC receiver with HD-PIC and SD-PIC branches.}
	\label{fig:pic_model}
\end{figure*}

The WL-MMSE equalizer jointly suppresses FTN-ISI, doubly selective channel spreading, and IQI image interference in the augmented domain. The resulting post equalization output and channel representation are used by the subsequent PIC stages. 
The residual off diagonal coupling in this representation mixes the decision inputs of different symbols and limits the reliability of direct symbol slicing after linear preprocessing. Therefore, HD-PIC and SD-PIC are introduced to further reconstruct and cancel this residual component through different feedback mechanisms, as summarized in Fig.~\ref{fig:pic_model}.

\subsection{Hard Decision PIC}
HD-PIC is introduced after WL-MMSE preprocessing to cancel the residual coupling that remains in the post equalization output. 
In the proposed HD-PIC stage, the current soft output is sliced into tentative hard decisions, and these decisions are used to reconstruct the interference associated with the off diagonal post equalization entries. 
The reconstructed residual term is then subtracted from the fixed WL-MMSE output to update the soft detection input. 

For subsequent residual interference reconstruction, the WL-MMSE output is first written in the post equalization form
\begin{equation}
	\scalebox{1}{$%
		\begin{aligned}
			\mathbf z^{(0)}=\mathbf W\mathbf y=\mathbf P\mathbf s+\mathbf v,
		\end{aligned}
		$}
	\label{eq:post_equal_model}
\end{equation}
where $\mathbf P=\mathbf W\mathcal H$ is the post equalization matrix and $\mathbf v=\mathbf W\mathbf n$ is the corresponding equivalent noise. Decompose $\mathbf P$ as $\mathbf P=\mathbf D+\mathbf P_{\rm off}$, where $\mathbf D=\diag(\mathbf P)$ contains the desired stream gains and $\mathbf P_{\rm off}$ contains the residual FTN-ISI, doubly selective channel spreading, and IQI image interference. The diagonal entries of $\mathbf D$ are assumed to be nonzero. In implementation, a small diagonal loading can be added during normalization when $|[\mathbf D]_{k,k}|$ is close to zero. The HD-PIC recursion then estimates and cancels the contribution of $\mathbf P_{\rm off}$ using tentative hard decisions.

The iteration is initialized with $\mathbf z^{(0)}$, i.e.,  $\mathbf z_h^{(0)}=\mathbf z^{(0)}$.
% The iteration is initialized by $\mathbf z_h^{(0)}=\mathbf z^{(0)}$. 
At the $i$th iteration, diagonal gain normalization removes the streamwise main gains from the current soft output, yielding
\begin{equation}
	\scalebox{1}{$%
		\begin{aligned}
\mathbf u_h^{(i)}
&=\mathbf D^{-1}\mathbf z_h^{(i)}.
		\end{aligned}
		$}
	\label{eq:hd_norm}
\end{equation}
Since $\mathbf u_h^{(i)}$ is an augmented vector, its first $N$ entries correspond to the original data symbols and are mapped to the nearest constellation points element by element. With $\hat{\mathbf d}^{(i)}=[\hat d_0^{(i)},\ldots,\hat d_{N-1}^{(i)}]^T$, the hard decision vector is determined by
\begin{equation}
	\scalebox{1}{$%
		\begin{aligned}
\hat d_k^{(i)}
&=\arg\min_{s\in\mathcal S}
\left|[\mathbf u_h^{(i)}]_k-s\right|^2,\quad k=0,\ldots,N-1.
		\end{aligned}
		$}
	\label{eq:hd_slice}
\end{equation}
The corresponding augmented tentative decision vector is constructed by
$
\hat{\mathbf s}^{(i)}=[(\hat{\mathbf d}^{(i)})^T \, (\hat{\mathbf d}^{(i)})^H]^T.
$
The estimated residual coupling term $\mathbf P_{\rm off}\hat{\mathbf s}^{(i)}$ is subtracted from the original WL-MMSE output, yielding the updated soft output
\begin{equation}
	\scalebox{1}{$%
		\begin{aligned}
\mathbf z_h^{(i+1)}=\mathbf z^{(0)}-\mathbf P_{\rm off}\hat{\mathbf s}^{(i)}.
		\end{aligned}
		$}
	\label{eq:hd_update}
\end{equation}
The update in \eqref{eq:hd_update} uses the same WL-MMSE output $\mathbf z^{(0)}$ and residual coupling matrix $\mathbf P_{\rm off}$ throughout the channel block, with $\hat{\mathbf s}^{(i)}$ providing the feedback at each iteration.

Let $I_{\max}$ denote the maximum number of PIC iterations and $\varepsilon_{\rm stop}$ denote the stopping threshold. 
The soft output convergence metric is defined as
\begin{equation}
\Delta_{\rm HD}^{(i+1)}= \frac{\|\mathbf u_h^{(i+1)}- \mathbf u_h^{(i)}\|_2^2} {\|\mathbf u_h^{(i)}\|_2^2}.
\label{eq:hd_delta}
\end{equation}
The HD-PIC iteration terminates once $\Delta_{\rm HD}^{(i+1)}\leq\varepsilon_{\rm stop}$ or it reaches the maximum number of iterations.
Let $I_{\rm HD,stop}$ denote the HD-PIC stopping iteration index. The final HD-PIC decision $\hat{\mathbf d}_{\rm HD}$ is obtained by applying the slicing rule in \eqref{eq:hd_slice} to $\mathbf u_h^{(I_{\rm HD,stop})}$.
Algorithm~\ref{alg:hd_pic} summarizes  the proposed  HD-PIC procedure.

\begin{algorithm}[t]
\caption{WL-MMSE Aided Hard Decision PIC}
\label{alg:hd_pic}
\begin{algorithmic}[1]
\REQUIRE $\mathbf y$, $\mathcal H$, $\mathbf R_n$, $\mathcal S$, $I_{\max}$, $\varepsilon_{\rm stop}$
\ENSURE $\hat{\mathbf d}_{\rm HD}$
\STATE Compute $\mathbf W$ from \eqref{eq:wlmmse}.
\STATE Set $\mathbf z^{(0)}=\mathbf W\mathbf y$ and $\mathbf P=\mathbf W\mathcal H$.
\STATE Set $\mathbf D=\diag(\mathbf P)$ and $\mathbf P_{\rm off}=\mathbf P-\mathbf D$.
\STATE Initialize the augmented output $\mathbf z_h^{(0)}=\mathbf z^{(0)}$ and set $I_{\rm HD,stop}=I_{\max}$.
\FOR{$i=0,1,\ldots,I_{\max}-1$}
\STATE Normalize the augmented output as $\mathbf u_h^{(i)}=\mathbf D^{-1}\mathbf z_h^{(i)}$.
\STATE $\hat d_k^{(i)}=\argmin\limits_{s\in\mathcal S}|[\mathbf u_h^{(i)}]_k-s|^2$, $k=0,\ldots,N-1$.
\STATE $\hat{\mathbf s}^{(i)}=[(\hat{\mathbf d}^{(i)})^T,(\hat{\mathbf d}^{(i)})^H]^T$.
\STATE $\mathbf z_h^{(i+1)}=\mathbf z^{(0)}-\mathbf P_{\rm off}\hat{\mathbf s}^{(i)}$.
\STATE $\mathbf u_h^{(i+1)}=\mathbf D^{-1}\mathbf z_h^{(i+1)}$.
\STATE Compute $\Delta_{\rm HD}^{(i+1)}=\|\mathbf u_h^{(i+1)}-\mathbf u_h^{(i)}\|_2^2/\|\mathbf u_h^{(i)}\|_2^2$ from the normalized outputs.
\IF{$\Delta_{\rm HD}^{(i+1)}\leq\varepsilon_{\rm stop}$}
\STATE Set $I_{\rm HD,stop}=i+1$.
\STATE \textbf{break}
\ENDIF
\ENDFOR
\STATE $\hat d_{{\rm HD},k}=\argmin\limits_{s\in\mathcal S}|[\mathbf u_h^{(I_{\rm HD,stop})}]_k-s|^2$, $k=0,\ldots,N-1$.
\end{algorithmic}
\end{algorithm}

Note that the proposed  HD-PIC receiver uses hard symbol estimates to reduce residual interference with an additional $O((2N)^2)$ matrix vector multiplication per iteration. The same residual cancellation principle can also incorporate reliability information from the WL-MMSE output. This leads to SD-PIC, which replaces hard feedback with posterior mean feedback and controls cancellation based on the variance and pseudo variance of the residual interference plus noise through a bivariate likelihood and damping.

\subsection{Soft Decision PIC}
This subsection presents the SD-PIC receiver under the same fixed residual coupling model used by HD-PIC. Instead of feeding back a single constellation point, SD-PIC uses posterior means together with per stream covariance and pseudo covariance statistics. Applying diagonal gain normalization to the WL-MMSE output gives the initial soft information as $\mathbf u^{(0)}=\mathbf D^{-1}\mathbf z^{(0)}$, and the corresponding normalized augmented noise covariance is given by
\begin{equation}
	\scalebox{1}{$%
		\begin{aligned}
			\mathbf R_v^{(0)}=\mathbf D^{-1}\mathbf W\mathbf R_n\mathbf W^H\mathbf D^{-H}.
		\end{aligned}
		$}
\label{eq:init_noise_cov}
\end{equation}
For data stream $k$, define the initial variance and pseudo variance as $\zeta_k^{(0)}=[\mathbf R_v^{(0)}]_{k,k}$ and $\chi_{v,k}^{(0)}=[\mathbf R_v^{(0)}]_{k,N+k}$, respectively. With $\Re\{\cdot\}$ and $\Im\{\cdot\}$ denoting the real and imaginary parts, the covariance of the real and imaginary residual components is
\begin{equation}
\scalebox{0.98}{$%
\begin{aligned}
\mathbf K_k^{(0)}=\frac{1}{2}
\begin{bmatrix}
\zeta_k^{(0)}+\Re\{\chi_{v,k}^{(0)}\} & \Im\{\chi_{v,k}^{(0)}\}\\
\Im\{\chi_{v,k}^{(0)}\} & \zeta_k^{(0)}-\Re\{\chi_{v,k}^{(0)}\}
\end{bmatrix}.
\end{aligned}
$}
\label{eq:init_real_cov}
\end{equation}
Following the second order interference and noise treatment in~\cite{xiao2011mimo_improved}, define $\mathbf r_k^{(0)}(s)=[\Re\{[\mathbf u^{(0)}]_k-s\},\Im\{[\mathbf u^{(0)}]_k-s\}]^T$. Using $\mathbf K_k^{(0)}$ as a noise based scale approximation, the initial posterior probability under a bivariate Student-$t$ type likelihood is given by~\cite{zoubir2012robust}
\begin{equation}
	\scalebox{0.98}{$%
		\begin{aligned}
		p_k^{(0)}(s)=\frac{\ell_k^{(0)}(s)}{\sum\limits_{s'\in\mathcal S}\ell_k^{(0)}(s')},\quad s\in\mathcal S,
		\end{aligned}
		$}
	\label{eq:init_prob}
\end{equation}
with
\begin{equation}
	\scalebox{0.98}{$%
		\begin{aligned}
			\ell_k^{(0)}(s)\triangleq\left(1+\frac{\mathbf r_k^{(0)}(s)^T[\mathbf K_k^{(0)}]^{-1}\mathbf r_k^{(0)}(s)}{\nu_s}\right)^{-(\nu_s+2)/2},
		\end{aligned}
		$}
	\label{eq:init_prob1}
\end{equation}
where $\nu_s>0$ controls the tail weight, with smaller values producing heavier tails and $\nu_s\rightarrow\infty$ recovering the Gaussian likelihood. The bivariate likelihood preserves the rotational variance of the improper residual instead of reducing it to a scalar uncertainty.
The resulting posterior mean, variance, and pseudo variance are respectively expressed as
\begin{subequations}
\label{eq:init_mean_var}
\begin{align}
\bar d_k^{(0)}&=\sum_{s\in\mathcal S}p_k^{(0)}(s)s,\label{eq:init_mean}\\
\xi_k^{(0)}&=\sum_{s\in\mathcal S}p_k^{(0)}(s)|s|^2-|\bar d_k^{(0)}|^2,\label{eq:init_var}\\
\psi_k^{(0)}&=\sum_{s\in\mathcal S}p_k^{(0)}(s)\left(s-\bar d_k^{(0)}\right)^2.\label{eq:init_pvar}
\end{align}
\end{subequations}
The initial posterior mean, variance, and pseudo variance vectors are $\bar{\mathbf d}^{(0)}=[\bar d_0^{(0)},\ldots,\bar d_{N-1}^{(0)}]^T$, $\boldsymbol\xi^{(0)}=[\xi_0^{(0)},\ldots,\xi_{N-1}^{(0)}]^T$, and $\boldsymbol\psi^{(0)}=[\psi_0^{(0)},\ldots,\psi_{N-1}^{(0)}]^T$, respectively.

At the $i$th iteration, the augmented mean vector is $\bar{\mathbf s}^{(i)}=[(\bar{\mathbf d}^{(i)})^T \, (\bar{\mathbf d}^{(i)})^H]^T$. The data branch covariance matrix is $\mathbf R_e^{(i)}=\diag\left(\xi_0^{(i)},\ldots,\xi_{N-1}^{(i)}\right)$, and its pseudo covariance matrix is $\mathbf C_e^{(i)}=\diag\left(\psi_0^{(i)},\ldots,\psi_{N-1}^{(i)}\right)$. The corresponding augmented error covariance is
 \begin{equation}
 \scalebox{0.90}{$%
 \begin{aligned}
 \boldsymbol\Sigma_e^{(i)}&=
 \begin{bmatrix}
 \mathbf R_e^{(i)} & \mathbf C_e^{(i)}\\
 \mathbf C_e^{(i)*} & \mathbf R_e^{(i)*}
 \end{bmatrix}.
 \end{aligned}
 $}
 \label{eq:soft_error_cov}
 \end{equation}
At iteration $i$, the current posterior mean reconstructs the residual coupling through $\mathbf P_{\rm off}$. The normalized soft output is given by
\begin{equation}
	\scalebox{1}{$%
		\begin{aligned}
\mathbf z_s^{(i+1)}
=\mathbf z^{(0)}-\mathbf P_{\rm off}\bar{\mathbf s}^{(i)},\quad
\mathbf u_s^{(i+1)}
=\mathbf D^{-1}\mathbf z_s^{(i+1)}.
		\end{aligned}
		$}
	\label{eq:soft_pic_update}
\end{equation}
The residual interference plus noise is generally improper after soft cancellation. Its augmented covariance at the PIC output is approximated by the following model:
\begin{equation}
\scalebox{0.98}{$%
\begin{aligned}
\boldsymbol\Sigma_r^{(i+1)}
 =\mathbf R_v^{(0)}
 +\boldsymbol\Gamma_{\rm res}\boldsymbol\Sigma_e^{(i)}\boldsymbol\Gamma_{\rm res}^H.
\end{aligned}
$}
\label{eq:soft_residual_stats}
\end{equation}
For each data stream $k=0,\ldots,N-1$, the effective variance and pseudo variance are extracted as
\begin{equation}
\scalebox{0.98}{$%
\begin{aligned}
\zeta_k^{(i+1)}=\left[\boldsymbol\Sigma_r^{(i+1)}\right]_{k,k},\quad
\chi_k^{(i+1)}=\left[\boldsymbol\Sigma_r^{(i+1)}\right]_{k,N+k}.
\end{aligned}
$}
\label{eq:soft_eff_stats}
\end{equation}
These entries can be evaluated by rowwise quadratic forms without explicitly storing $\boldsymbol\Sigma_r^{(i+1)}$, with the same $\mathcal O((2N)^2)$ order as the scalar evaluation. The real valued covariance associated with $\zeta_k^{(i+1)}$ and $\chi_k^{(i+1)}$ is
\begin{equation}
\scalebox{0.93}{$%
\begin{aligned}
\mathbf K_k^{(i+1)}=\frac{1}{2}
\begin{bmatrix}
\zeta_k^{(i+1)}+\Re\{\chi_k^{(i+1)}\} & \Im\{\chi_k^{(i+1)}\}\\
\Im\{\chi_k^{(i+1)}\} & \zeta_k^{(i+1)}-\Re\{\chi_k^{(i+1)}\}
\end{bmatrix}.
\end{aligned}
$}
\label{eq:soft_real_cov}
\end{equation}
With $\mathbf r_k^{(i+1)}(s)=[\Re\{[\mathbf u_s^{(i+1)}]_k-s\},\Im\{[\mathbf u_s^{(i+1)}]_k-s\}]^T$, the posterior probabilities are updated according to the same bivariate Student-$t$ type likelihood, with $\mathbf K_k^{(i+1)}$ used as the scale matrix
\begin{equation}
	\scalebox{0.98}{$%
		\begin{aligned}
			p_k^{(i+1)}(s)=\frac{\ell_k^{(i+1)}(s)}{\sum\limits_{s'\in\mathcal S}\ell_k^{(i+1)}(s')},\quad s\in\mathcal S,
		\end{aligned}
		$}
	\label{eq:sd_prob_update}
\end{equation}
with
\begin{equation}
	\scalebox{0.98}{$%
		\begin{aligned}
			\ell_k^{(i+1)}(s)\triangleq\left(1+\frac{\mathbf r_k^{(i+1)}(s)^T[\mathbf K_k^{(i+1)}]^{-1}\mathbf r_k^{(i+1)}(s)}{\nu_s}\right)^{-(\nu_s+2)/2}.
		\end{aligned}
		$}
	\label{eq:sd_prob_update1}
\end{equation}
From the first and second moments of $p_k^{(i+1)}(s)$, the target posterior mean, variance, and pseudo variance are
\begin{subequations}
\label{eq:target_mean_var}
\begin{align}
\bar d_{{\rm tar},k}^{(i+1)}&=\sum_{s\in\mathcal S}p_k^{(i+1)}(s)s,\label{eq:target_mean}\\
\xi_{{\rm tar},k}^{(i+1)}&=\sum_{s\in\mathcal S}p_k^{(i+1)}(s)|s|^2-|\bar d_{{\rm tar},k}^{(i+1)}|^2,\label{eq:target_var}\\
\psi_{{\rm tar},k}^{(i+1)}&=\sum_{s\in\mathcal S}p_k^{(i+1)}(s)\left(s-\bar d_{{\rm tar},k}^{(i+1)}\right)^2.\label{eq:target_pvar}
\end{align}
\end{subequations}
After collecting the target entries into $\bar{\mathbf d}_{\rm tar}^{(i+1)}$, $\boldsymbol\xi_{\rm tar}^{(i+1)}$, and $\boldsymbol\psi_{\rm tar}^{(i+1)}$, damping is introduced to prevent unreliable posterior updates from causing abrupt changes in the soft feedback. The posterior mean, variance, and pseudo variance are updated as
\begin{subequations}
\label{eq:mean_var_update}
\begin{align}
\bar{\mathbf d}^{(i+1)}&=(1-\eta_i)\bar{\mathbf d}^{(i)}+\eta_i\bar{\mathbf d}_{\rm tar}^{(i+1)},
\label{eq:mean_update}\\
\boldsymbol\xi^{(i+1)}&=(1-\eta_i)\boldsymbol\xi^{(i)}+\eta_i\boldsymbol\xi_{\rm tar}^{(i+1)},
\label{eq:var_update}
\\
\boldsymbol\psi^{(i+1)}&=(1-\eta_i)\boldsymbol\psi^{(i)}+\eta_i\boldsymbol\psi_{\rm tar}^{(i+1)},
\label{eq:pvar_update}
\end{align}
\end{subequations}
where $\eta_i$ denotes the damping factor at the $i$th iteration, initialized by $\eta_0$ and constrained by $0<\eta_{\min}\leq\eta_i\leq1$.
The variance entries are clipped to $[\xi_{\min},\xi_{\max}]$ to avoid overconfident feedback, and the pseudo variance is projected onto the admissible set $|\psi_k^{(i+1)}|\leq\xi_k^{(i+1)}$ to preserve a valid second order statistic. Here, $\xi_{\min}$ and $\xi_{\max}$ denote the lower and upper clipping bounds, respectively. The empirical reliability cost used to adjust the damping factor is given by
\begin{equation}
	\scalebox{1}{$%
		\begin{aligned}
J^{(i+1)}=\frac{\|\mathbf y-\mathcal H\bar{\mathbf s}^{(i+1)}\|_2^2}{\|\mathbf y\|_2^2}+\lambda\frac{1}{N}\sum_{k=0}^{N-1}\xi_k^{(i+1)},
		\end{aligned}
		$}
	\label{eq:soft_cost}
\end{equation}
where $\lambda\geq0$ sets the weight of the posterior variance penalty relative to the normalized residual term and thereby influences the acceptance of trial updates during backtracking.
Let $J_{\rm best}$ denote the previously accepted reliability cost. If a trial update gives $J^{(i+1)}>J_{\rm best}$, the damping factor is reduced by $\gamma_{\rm dec}\in(0,1)$ and the update is recomputed until $J^{(i+1)}\leq J_{\rm best}$ or $\eta_i=\eta_{\min}$. Variance clipping and pseudo variance projection are repeated after each backtracking update. If no acceptable decrease is obtained at $\eta_{\min}$, the previous posterior statistics are retained and the stopping test is skipped. For an accepted update, the next damping factor is increased by $\gamma_{\rm inc}>1$ and capped by one. Let $I_{\rm SD,stop}\leq I_{\max}$ denote the stopping index, determined by the maximum iteration number or by the posterior mean convergence criterion $\Delta_{\rm SD}^{(i+1)}\leq\varepsilon_{\rm stop}$, where $\Delta_{\rm SD}^{(i+1)}=\|\bar{\mathbf d}^{(i+1)}-\bar{\mathbf d}^{(i)}\|_2^2/\|\bar{\mathbf d}^{(i)}\|_2^2$. With $\hat{\mathbf d}_{\rm SD}=[\hat d_{{\rm SD},0},\ldots,\hat d_{{\rm SD},N-1}]^T$, the final data decision is
\begin{equation}
	\scalebox{0.98}{$%
		\begin{aligned}
\hat d_{{\rm SD},k}
=\arg\min_{s\in\mathcal S}
\left|\bar d_k^{(I_{\rm SD,stop})}-s\right|^2,\quad k=0,\ldots,N-1.
		\end{aligned}
		$}
	\label{eq:sd_final}
\end{equation}
Algorithm~\ref{alg:sd_pic} summarizes the SD-PIC procedure.
\begin{algorithm}[t]
\caption{WL-MMSE Aided Soft Decision PIC}
\label{alg:sd_pic}
\begin{algorithmic}[1]
\REQUIRE $\mathbf y$, $\mathcal H$, $\mathbf R_n$, $\mathcal S$, $I_{\max}$, $\nu_s$, $\lambda$, $0<\eta_{\min}\leq\eta_0\leq1$, $\gamma_{\rm dec}$, $\gamma_{\rm inc}$, $\xi_{\min}$, $\xi_{\max}$, $\varepsilon_{\rm stop}$
\ENSURE $\hat{\mathbf d}_{\rm SD}$
\STATE Compute $\mathbf W$ from \eqref{eq:wlmmse}. Set $\mathbf z^{(0)}=\mathbf W\mathbf y$, $\mathbf P=\mathbf W\mathcal H$, $\mathbf D=\diag(\mathbf P)$, $\mathbf P_{\rm off}=\mathbf P-\mathbf D$, and $\boldsymbol\Gamma_{\rm res}=\mathbf D^{-1}\mathbf P_{\rm off}$.
\STATE Compute $\mathbf u^{(0)}=\mathbf D^{-1}\mathbf z^{(0)}$ and evaluate $\mathbf R_v^{(0)}$ using \eqref{eq:init_noise_cov}.
\STATE Compute $p_k^{(0)}(s)$ from $\mathbf u^{(0)}$ and $\mathbf R_v^{(0)}$ using \eqref{eq:init_real_cov}--\eqref{eq:init_prob1}.
\STATE Initialize the data domain moments $\bar{\mathbf d}^{(0)}$, $\boldsymbol\xi^{(0)}$, and $\boldsymbol\psi^{(0)}$ using \eqref{eq:init_mean_var}.
\STATE Form $\bar{\mathbf s}^{(0)}$ and evaluate $J^{(0)}$ using \eqref{eq:soft_cost}. Set $\eta=\eta_0$, $J_{\rm best}=J^{(0)}$, and $I_{\rm SD,stop}=I_{\max}$.
\FOR{$i=0,1,\ldots,I_{\max}-1$}
\STATE Form $\bar{\mathbf s}^{(i)}$ and the augmented error covariance $\boldsymbol\Sigma_e^{(i)}$ from the current data domain moments using \eqref{eq:soft_error_cov}.
\STATE Compute $\mathbf z_s^{(i+1)}$ and $\mathbf u_s^{(i+1)}$ using \eqref{eq:soft_pic_update}.
\STATE Evaluate $\zeta_k^{(i+1)}$ and $\chi_k^{(i+1)}$ rowwise from \eqref{eq:soft_residual_stats} and \eqref{eq:soft_eff_stats}.
\STATE Compute $p_k^{(i+1)}(s)$, $\bar{\mathbf d}_{\rm tar}^{(i+1)}$, $\boldsymbol\xi_{\rm tar}^{(i+1)}$, and $\boldsymbol\psi_{\rm tar}^{(i+1)}$.
\STATE Set $\eta_i=\eta$ and compute trial data domain moments $\bar{\mathbf d}^{(i+1)}$, $\boldsymbol\xi^{(i+1)}$, and $\boldsymbol\psi^{(i+1)}$ using \eqref{eq:mean_var_update}.
\STATE Clip $\boldsymbol\xi^{(i+1)}$ to $[\xi_{\min},\xi_{\max}]$.
\STATE Project each $\psi_k^{(i+1)}$ onto $|\psi_k^{(i+1)}|\leq\xi_k^{(i+1)}$.
\STATE Evaluate $J^{(i+1)}$ using \eqref{eq:soft_cost}.
\WHILE{$J^{(i+1)}>J_{\rm best}$ and $\eta_i>\eta_{\min}$}
\STATE Set $\eta_i=\max(\eta_{\min},\gamma_{\rm dec}\eta_i)$.
\STATE Recompute $\bar{\mathbf d}^{(i+1)}$, $\boldsymbol\xi^{(i+1)}$, and $\boldsymbol\psi^{(i+1)}$ using \eqref{eq:mean_var_update}.
\STATE Clip $\boldsymbol\xi^{(i+1)}$ to $[\xi_{\min},\xi_{\max}]$.
\STATE Project each $\psi_k^{(i+1)}$ onto $|\psi_k^{(i+1)}|\leq\xi_k^{(i+1)}$.
\STATE Recompute $J^{(i+1)}$ using \eqref{eq:soft_cost}.
\ENDWHILE
\IF{$J^{(i+1)}\leq J_{\rm best}$}
\STATE Accept the update, set $J_{\rm best}=J^{(i+1)}$, and set $\eta=\min(1,\gamma_{\rm inc}\eta_i)$.
\STATE Compute $\Delta_{\rm SD}^{(i+1)}=\|\bar{\mathbf d}^{(i+1)}-\bar{\mathbf d}^{(i)}\|_2^2/\|\bar{\mathbf d}^{(i)}\|_2^2$ from the accepted data domain means.
\IF{$\Delta_{\rm SD}^{(i+1)}\leq\varepsilon_{\rm stop}$}
\STATE Set $I_{\rm SD,stop}=i+1$.
\STATE \textbf{break}
\ENDIF
\ELSE
\STATE Retain $\bar{\mathbf d}^{(i+1)}=\bar{\mathbf d}^{(i)}$, $\boldsymbol\xi^{(i+1)}=\boldsymbol\xi^{(i)}$, and $\boldsymbol\psi^{(i+1)}=\boldsymbol\psi^{(i)}$.
\STATE Set $\Delta_{\rm SD}^{(i+1)}=\infty$ and skip the stopping test.
\ENDIF
\ENDFOR
\STATE $\hat d_{{\rm SD},k}=\argmin\limits_{s\in\mathcal S}|\bar d_k^{(I_{\rm SD,stop})}-s|^2$, $k=0,\ldots,N-1$.
\end{algorithmic}
\end{algorithm}
\subsection{Computational Complexity}
The receiver complexity is evaluated using the dominant matrix operations, as commonly adopted for linear and iterative receivers~\cite{li2020ftn_equalization}. Conventional L-MMSE operates on the model without augmentation and is dominated by an $N\times N$ matrix inversion, yielding $\mathcal O(N^3)$. WL-MMSE processes the real and image components in the augmented domain, which requires a $2N\times 2N$ matrix inversion together with post equalization matrix computation and therefore leads to a dominant complexity of $\mathcal O((2N)^3)$.

For WL-MMSE-HD-PIC, WL-MMSE preprocessing and the decomposition of $\mathbf P$ are computed before the iterative cancellation stage and kept fixed during the subsequent iterations. Let $I_{\rm HD}$ and $I_{\rm SD}$ denote the numbers of HD-PIC and SD-PIC iterations, respectively. The HD-PIC iterations are dominated by diagonal gain normalization, symbolwise slicing, and residual interference reconstruction through $\mathbf P_{\rm off}\hat{\mathbf s}^{(i)}$. Accordingly, the additional iterative complexity relative to WL-MMSE is $I_{\rm HD}\mathcal O((2N)^2)$. For WL-MMSE-SD-PIC, the WL-MMSE filter is also computed once, and each iteration requires soft residual reconstruction, variance and pseudo variance evaluation, bivariate posterior probability calculation, and second order statistic updates. Evaluating the bivariate likelihood and updating the posterior mean, variance, and pseudo variance require $\mathcal O(NM)$ operations, while the rowwise quadratic forms for residual variance and pseudo variance require $\mathcal O((2N)^2)$ operations. Since the full residual covariance matrix is not explicitly formed, the additional iterative complexity remains $I_{\rm SD}[\mathcal O((2N)^2)+\mathcal O(NM)]$. The AFDM chirp parameter search is performed separately and is excluded from the per block receiver complexity once $c_1$ has been selected. For $|\mathcal Q_c|$ candidates, its cost is approximately $|\mathcal Q_c|$ times the sum of candidate preprocessing and residual leakage evaluation costs. The resulting complexity comparison is summarized in Table~\ref{tab:complexity}.

\begin{table}[t]
\caption{Complexity Analysis for Different Algorithms}
\label{tab:complexity}
\centering
\scriptsize
\renewcommand{\arraystretch}{1.2}
\begin{tabular}{c|c}
\hline
\textbf{Algorithm} & \textbf{Complexity} \\ \hline \hline
L-MMSE & $\mathcal O(N^3)$ \\ \hline
WL-MMSE & $\mathcal O((2N)^3)$ \\ \hline
WL-MMSE-HD-PIC & $\mathcal O((2N)^3)+I_{\rm HD}\mathcal O((2N)^2)$ \\ \hline
WL-MMSE-SD-PIC & $\mathcal O((2N)^3)+I_{\rm SD}\left[\mathcal O((2N)^2)+\mathcal O(NM)\right]$ \\ \hline
\end{tabular}
\end{table}

%%=======================================Chapter 4 Theoretical Analysis=======================================%%
\section{Residual Coupling Based Receiver Analysis and Parameter Selection}
\label{sec:4}
This section analyzes residual coupling stability, second order soft feedback reliability, genie aided performance, and receiver oriented AFDM-FTN parameter selection. The SD-PIC analysis accounts for posterior variances, pseudo variances, and damping control.

\subsection{Local Stability Analysis of PIC}
This subsection provides a local stability interpretation of the PIC recursion. The analysis characterizes how the residual coupling matrix propagates feedback errors after WL-MMSE preprocessing. Based on the post equalization model in \eqref{eq:post_equal_model}, substituting the HD-PIC update in \eqref{eq:hd_update} yields the soft output at the $(i+1)$th iteration,
\begin{equation}
	\scalebox{1}{$%
		\begin{aligned}
\mathbf z_h^{(i+1)}=\mathbf D\mathbf s+\mathbf P_{\rm off}(\mathbf s-\hat{\mathbf s}^{(i)})+\mathbf v.
		\end{aligned}
		$}
	\label{eq:hd_convergence_output}
\end{equation}
To express \eqref{eq:hd_convergence_output} in a gain normalized form, the residual coupling matrix and equivalent noise are defined with respect to $\mathbf D$ as
\begin{equation}
	\scalebox{1}{$%
		\begin{aligned}
\boldsymbol\Gamma_{\rm res}=\mathbf D^{-1}\mathbf P_{\rm off},
\quad
\bar{\mathbf v}=\mathbf D^{-1}\mathbf v.
		\end{aligned}
		$}
	\label{eq:normalized_coupling}
\end{equation}
Premultiplying \eqref{eq:hd_convergence_output} by $\mathbf D^{-1}$ and using \eqref{eq:normalized_coupling} gives the normalized HD-PIC recursion
\begin{equation}
	\scalebox{1}{$%
		\begin{aligned}
\mathbf u_h^{(i+1)}=\mathbf s+\boldsymbol\Gamma_{\rm res}\mathbf e_h^{(i)}+\bar{\mathbf v},
		\end{aligned}
		$}
	\label{eq:hd_norm_recursion}
\end{equation}
where $\mathbf e_h^{(i)}=\mathbf s-\hat{\mathbf s}^{(i)}$ is the hard decision error. \eqref{eq:hd_norm_recursion} separates the desired symbol, the propagated decision error, and the normalized noise. Hence, the influence of an erroneous tentative decision is governed by $\boldsymbol\Gamma_{\rm res}$, and a smaller $\|\boldsymbol\Gamma_{\rm res}\|_\infty$ corresponds to stronger post equalization diagonal dominance.

Denote $\delta_{\min}$ as the minimum Euclidean distance from the transmitted constellation point to its nearest decision boundary. 
According to \eqref{eq:hd_norm_recursion}, the entries indexed by $0,\ldots,N-1$ of the slicing input correspond to the data symbol vector $\mathbf d$ and contain the aggregate perturbation $[\boldsymbol\Gamma_{\rm res}\mathbf e_h^{(i)}+\bar{\mathbf v}]_{0:N-1}$.
A sufficient condition for preserving the correct decision region at the $i$th iteration is
\begin{equation}
	\scalebox{1}{$%
		\begin{aligned}
            \left\|[\boldsymbol\Gamma_{\rm res}\mathbf e_h^{(i)}+\bar{\mathbf v}]_{0:N-1}\right\|_\infty<\delta_{\min}.
		\end{aligned}
		$}
	\label{eq:correct_condition}
\end{equation}
Under \eqref{eq:correct_condition}, the slicing input $\mathbf u_h^{(i+1)}$ remains inside the correct Voronoi decision region, yielding a correct hard decision at the next iteration. Once this state is reached, the decision error term vanishes, and the state is preserved if the normalized noise satisfies $\left\|[\bar{\mathbf v}]_{0:N-1}\right\|_\infty<\delta_{\min}$.
Thus, condition \eqref{eq:correct_condition} is a one step criterion for finite alphabet slicing in HD-PIC.
For analytical comparison with HD-PIC, SD-PIC can be interpreted under the fixed post equalization residual coupling approximation, where hard feedback is replaced by posterior mean feedback. Define the soft feedback error by
\begin{equation}
	\scalebox{1}{$%
		\begin{aligned}
\mathbf e_s^{(i)}
=\mathbf s-\bar{\mathbf s}^{(i)}=\left[
(\mathbf d-\bar{\mathbf d}^{(i)})^T \,
(\mathbf d-\bar{\mathbf d}^{(i)})^H
\right]^T.
		\end{aligned}
		$}
	\label{eq:soft_error}
\end{equation}
The corresponding normalized residual perturbation follows the same form as \eqref{eq:hd_norm_recursion}, namely
\begin{equation}
	\scalebox{1}{$%
		\begin{aligned}
            \mathbf u_s^{(i+1)}=\mathbf s+\boldsymbol\Gamma_{\rm res}\mathbf e_s^{(i)}+\bar{\mathbf v}.
		\end{aligned}
		$}
	\label{eq:sd_norm_recursion}
\end{equation}
Unlike the discrete hard decision error in HD-PIC, the SD-PIC feedback error is controlled by posterior second order statistics. For the $k$th data stream and $i\geq1$, the target posterior variance and pseudo variance before damping and clipping are given by
\begin{equation}
	\scalebox{1}{$%
		\begin{aligned}
            {\xi_{{\rm tar},k}^{(i)}
            =\sum\limits_{s\in\mathcal S}p_k^{(i)}(s)|s|^2
            -\left|\bar d_{{\rm tar},k}^{(i)}\right|^2,}\\
            {\psi_{{\rm tar},k}^{(i)}
            =\sum\limits_{s\in\mathcal S}p_k^{(i)}(s)
            \left(s-\bar d_{{\rm tar},k}^{(i)}\right)^2.}
		\end{aligned}
		$}
	\label{eq:posterior_var_interp}
\end{equation}
A larger $\xi_{{\rm tar},k}^{(i)}$ indicates greater uncertainty in the target posterior mean, while a pseudo variance magnitude close to $\xi_{{\rm tar},k}^{(i)}$ indicates stronger impropriety of the target feedback error. The damped, clipped, and projected feedback statistics form $\boldsymbol\Sigma_e^{(i)}$, used to approximate $\boldsymbol\Sigma_{e_s}^{(i)}$. These statistics are subsequently mixed across data streams by $\boldsymbol\Gamma_{\rm res}$. Consequently, the covariance and pseudo covariance of the propagated residual error cannot in general be represented by their diagonal entries alone. Under the fixed residual coupling model, the resulting augmented second order statistics are given by~\cite{xiao2011mimo_improved}
\begin{equation}
	\scalebox{0.98}{$%
		\begin{aligned}
\boldsymbol\Sigma_{\rm prop}^{(i+1)}
&=\boldsymbol\Gamma_{\rm res}
\boldsymbol\Sigma_{e_s}^{(i)}
\boldsymbol\Gamma_{\rm res}^{H},\quad
\boldsymbol\Sigma_{e_s}^{(i)}
=\mathbb E\!\left[\mathbf e_s^{(i)}
(\mathbf e_s^{(i)})^H\mid\mathbf y\right],
		\end{aligned}
		$}
	\label{eq:full_cov_propagation}
\end{equation}
The approximation $\boldsymbol\Sigma_{e_s}^{(i)}\approx\boldsymbol\Sigma_e^{(i)}$ uses the augmented structure in \eqref{eq:soft_error_cov}. Including the normalized noise and neglecting its cross moments with the feedback error gives $\boldsymbol\Sigma_{\rm out}^{(i+1)}\approx\boldsymbol\Sigma_{\rm prop}^{(i+1)}+\mathbf R_v^{(0)}$, consistent with \eqref{eq:soft_residual_stats}. The additive noise determines the local error floor but does not change the homogeneous covariance contraction condition. Under fixed residual coupling, the vectorized propagation is expressed as
\begin{equation}
	\scalebox{0.98}{$%
		\begin{aligned}
\operatorname{vec}\!\left(\boldsymbol\Sigma_{\rm prop}^{(i+1)}\right)
&=\left(\boldsymbol\Gamma_{\rm res}^{*}\otimes
\boldsymbol\Gamma_{\rm res}\right)
\operatorname{vec}\!\left(\boldsymbol\Sigma_{e_s}^{(i)}\right).
		\end{aligned}
		$}
	\label{eq:covariance_stability}
\end{equation}
Accordingly, a stability condition for the homogeneous second moment propagation is given by
\begin{equation}
	\scalebox{1}{$%
		\begin{aligned}
\rho\!\left(\boldsymbol\Gamma_{\rm res}^{*}\otimes
\boldsymbol\Gamma_{\rm res}\right)<1,
\end{aligned}
		$}
	\label{eq:full_cov_stability}
\end{equation}
where $\rho(\cdot)$ denotes the spectral radius. The simpler condition $\|\boldsymbol\Gamma_{\rm res}\|_2<1$ is sufficient for \eqref{eq:full_cov_stability}.

Condition \eqref{eq:full_cov_stability} characterizes contraction of the homogeneous covariance and pseudo covariance propagation under fixed residual coupling. Posterior updating provides the soft estimates and their second order statistics, variance clipping and pseudo variance projection maintain admissible feedback statistics, and damping regulates the update step.

\subsection{Genie Aided Analytical Reference}
A genie aided analytical reference is introduced to characterize the proposed PIC receiver under perfect residual coupling cancellation. The genie aided PIC uses the true transmitted symbols to reconstruct and cancel residual coupling exactly. The GAB starts from this canceled state and models each stream using a scalar equivalent SNR and a proper Gaussian noise approximation.

According to \eqref{eq:post_equal_model} and $\mathbf P=\mathbf D+\mathbf P_{\rm off}$, perfect residual coupling cancellation leaves only the diagonal gain component and the post equalization noise. After diagonal normalization, the genie aided output becomes $\mathbf z_\infty=\mathbf s+\breve{\mathbf v}$ with $\breve{\mathbf v}=\mathbf D^{-1}\mathbf W\mathbf n$. Hence, its covariance matrix is identical to the normalized equivalent noise covariance in \eqref{eq:init_noise_cov}, i.e., $\mathbf R_{\breve v}=\mathbf R_v^{(0)}$. For the $k$th original data stream, the corresponding scalar equivalent output signal to noise ratio (SNR) can be written as
\begin{equation}
	\scalebox{1}{$%
		\begin{aligned}
            \gamma_{0,k}
            &=\frac{\mathbb E[|d_k|^2]}{\mathbb E[|\breve v_k|^2]}
            =\frac{\sigma_d^2}{[\mathbf R_{\breve v}]_{k,k}}  \\
            &=\frac{\sigma_d^2}
            {[\mathbf D^{-1}\mathbf W\mathbf R_n\mathbf W^H\mathbf D^{-H}]_{k,k}},
            \, k=0,\ldots,N-1.
		\end{aligned}
		$}
	\label{eq:gab_snr}
\end{equation}
Under the scalar equivalent SNR approximation, the normalized post equalization noise of each stream is treated as a proper complex Gaussian random variable with variance $[\mathbf R_{\breve v}]_{k,k}$. The symbol error rate of the $k$th stream for square $M$-QAM follows the standard signal space error probability approximation~\cite{lu1999qam_ber},
\begin{equation}
	\scalebox{1}{$%
		\begin{aligned}
            P_{s,k}^{\rm GAB}\approx 1-\left[1-2\left(1-\frac{1}{\sqrt M}\right)Q\left(\sqrt{\frac{3\gamma_{0,k}}{M-1}}\right)\right]^2,
		\end{aligned}
		$}
	\label{eq:gab_ser}
\end{equation}
where the standard $Q$-function is defined as
\begin{equation}
	\scalebox{1}{$%
		\begin{aligned}
            Q(x)=\frac{1}{\sqrt{2\pi}}\int_x^\infty e^{-t^2/2}dt.
		\end{aligned}
		$}
	\label{eq:q_function}
\end{equation}
With the Gray mapping approximation, the average BER corresponding to the genie aided reference can be evaluated by
\begin{equation}
	\scalebox{1}{$%
		\begin{aligned}
            P_b^{\rm GAB}\approx \frac{1}{N\log_2 M}\sum_{k=0}^{N-1}P_{s,k}^{\rm GAB}.
		\end{aligned}
		$}
	\label{eq:gab_ber}
\end{equation}
This reference is common to HD-PIC and SD-PIC because both receivers reduce to the same ideal residual coupling cancellation state under perfect side information.

% A gap between practical BER and the genie-aided reference mainly results from incomplete residual coupling reconstruction, tentative decision error propagation, biased posterior mean and variance estimation, convergence loss caused by damping, and the insufficiency of scalar equivalent SNR approximation in characterizing correlated noise and finite frame fluctuations. In simulations, a genie-aided PIC curve that uses the true transmitted symbols for residual reconstruction can be included to quantify the distance from practical receivers to the ideal cancellation state.

\subsection{Residual Leakage Based Parameter Selection}

The local stability analysis and the genie aided analytical reference both indicate that the receiver performance is governed by the residual coupling after WL-MMSE preprocessing. This coupling is affected by two controllable parameters: the AFDM chirp parameter $c_1$, which shapes the delay-Doppler-induced interference distribution in the DAFT domain, and the FTN compression factor $\alpha$, which determines the strength of FTN-ISI.
The chirp parameter $c_1$ is selected on the discrete AFDM grid according to
\begin{equation}
	\scalebox{1}{$%
		\begin{aligned}
            c_1=\frac{q}{2N},
		\end{aligned}
		$}
\end{equation}
where $q$ is an integer satisfying the AFDM parameter constraint. For a given $\alpha$, channel condition, SNR, and IQI level, $q$ is selected according to the residual coupling distribution induced by DAFT domain spreading. For each pair $(q,\alpha)$, let $\mathbf P^{(q,\alpha)}$ denote the post equalization matrix and define
\begin{equation}
	\scalebox{1}{$%
		\begin{aligned}
			\mathbf P_{\rm off}^{(q,\alpha)} \triangleq \mathbf P^{(q,\alpha)}-\diag(\mathbf P^{(q,\alpha)}).
		\end{aligned}
		$}
	\label{eq:param_poff}
\end{equation}
Since final slicing is performed on the first $N$ entries of the augmented vector, the residual leakage metric is evaluated on the data branch submatrix
\begin{equation}
	\scalebox{1}{$%
		\begin{aligned}
            \mathbf P_{{\rm off},d}^{(q,\alpha)}
            \triangleq
            \left[\mathbf P_{\rm off}^{(q,\alpha)}\right]_{0:N-1,0:N-1}.
		\end{aligned}
		$}
	\label{eq:param_poff_data}
\end{equation}
For a circular offset $o$, let $\mathcal I_o\triangleq\{(m,n):[m-n]_N=o,\ 0\leq m,n\leq N-1,\ m\neq n\}$. The corresponding residual energy is given by
\begin{equation}
	\scalebox{1}{$%
		\begin{aligned}
            E_o(q,\alpha)=
            \sum_{(m,n)\in\mathcal I_o}
            \left|[\mathbf P_{{\rm off},d}^{(q,\alpha)}]_{m,n}\right|^2.
		\end{aligned}
		$}
	\label{eq:offset_energy}
\end{equation}
The final chirp index is selected by minimizing the weighted residual leakage metric over the feasible chirp grid, namely
\begin{equation}
	\scalebox{1}{$%
		\begin{aligned}
            q^*=\arg\min_{q\in\mathcal Q_c}J_c(q;\alpha),
		\end{aligned}
		$}
\end{equation}
where $\mathcal Q_c$ denotes the feasible chirp grid index set. The metric $J_c(q;\alpha)$ is defined by
\begin{equation}
	\scalebox{1}{$%
		\begin{aligned}
            J_c(q;\alpha)=\sum_{o\in\Omega}E_o(q,\alpha)\omega(d_o),
		\end{aligned}
		$}
\end{equation}
where $\Omega$ denotes the circular offset set considered in the residual leakage metric, $d_o=\min(o,N-o)$ is the cyclic distance from offset $o$ to the main diagonal, and $\omega(d)=1/(d+1)$ assigns larger weights to residual components closer to the decision relevant diagonal entries.

Let $q_{\rm ref}\in\mathcal Q_c$ denote a reference chirp grid index. The corresponding reference residual offset energy is $E_o^{\rm ref}(\alpha)\triangleq E_o(q_{\rm ref},\alpha)$.
% \begin{equation}
% 	\scalebox{1}{$%
% 		\begin{aligned}
%             E_o^{\rm ref}(\alpha)\triangleq E_o(q_{\rm ref},\alpha).
% 		\end{aligned}
% 		$}
% 	\label{eq:offset_energy_ref}
% \end{equation}
For a prescribed residual energy coverage ratio $\kappa_E\in(0,1)$, the reference effective residual width $D_{\rm eff}(\alpha)$ is the smallest integer satisfying
\begin{equation}
	\scalebox{1}{$%
		\begin{aligned}
            \frac{\sum_{o\in\Omega:\,\min(o,N-o)\leq D_{\rm eff}(\alpha)}E_o^{\rm ref}(\alpha)}
            {\sum_{o\in\Omega}E_o^{\rm ref}(\alpha)}
            \geq \kappa_E.
		\end{aligned}
		$}
	\label{eq:deff}
\end{equation}
Based on this reference effective width, a heuristic initial chirp grid index is
\begin{equation}
	\scalebox{1}{$%
		\begin{aligned}
            q_0=\left\lfloor\frac{N}{D_{\rm eff}(\alpha)+1}\right\rfloor.
		\end{aligned}
		$}
\end{equation}
The corresponding initial chirp parameter is set to $c_{1,0}=q_0/2N$. 
Compared with the conventional AFDM rule based on the maximum delay and Doppler spread, the reference profile can incorporate FTN pulse correlation and IQI induced image leakage. The value $q_0$ is used only as a heuristic initialization or search center candidate for reducing the chirp grid search range. The final chirp index is still selected by evaluating $J_c(q;\alpha)$ over the feasible grid or by BER validation.

The FTN compression factor $\alpha$ controls the tradeoff between the spectral efficiency contribution and FTN-ISI~\cite{rusek2009capacity,anderson2013ftn}. A smaller $\alpha$ increases the rate contribution but also strengthens the post equalization residual coupling.
For a prescribed BER or error floor requirement, the admissible compression factor is determined by the residual leakage constraint, i.e.,
\begin{equation}
	\scalebox{1}{$%
		\begin{aligned}
			\alpha^*=\min_{\alpha\in\mathcal A}\alpha, \quad {\rm s.t.}\quad \min_{q\in\mathcal Q_c}J_c(q;\alpha)\leq J_{\rm th},
		\end{aligned}
		$}
\end{equation}
where $J_{\rm th}$ is the prescribed residual leakage threshold. Thus, $\alpha$ specifies the admissible interference level associated with FTN compression, while $c_1$ is selected on the AFDM chirp grid to reduce the corresponding residual leakage. The design procedure is therefore to determine the feasible range of $\alpha$ from the target BER and rate requirements, and then choose $c_1$ according to the residual leakage criterion.

%%=======================================Chapter 5 Simulation Results and Discussion ============================================%%
\begin{table}[t]
\centering
\caption{Main Simulation Parameters}
\label{tab:sim_params}
\scriptsize
\renewcommand{\arraystretch}{1.2}
\begin{tabular}{c|c}
\hline
\textbf{Parameter} & \textbf{Value} \\ \hline \hline
Modulation & 4-QAM (default), 16-QAM \\ \hline
AFDM block  size & $N=256$ \\ \hline
CPP length & $L_{\rm cpp}=10$ \\ \hline
Number of channel paths & $P=3$ \\ \hline
% Delay taps & $\{0,1,2\}$ \\ \hline
% Doppler bins & Random in $\{-1,0,1\}$ \\ \hline
%Channel power profile & Normalized equal-power paths \\ \hline
RRC rolloff factor & $\beta=0.3$ \\ \hline
Pulse half span & $K_g=9$ \\ \hline
%Pulse span & $20$ symbols \\ \hline
% SNR range & $1:1:20$ dB \\ \hline
% Monte Carlo frames & $5\times 10^3$ per SNR \\ \hline
AFDM chirp parameter $c_1$ & $3/(2N)$ \\ \hline
AFDM chirp parameter $c_2$ & $1/(2N^2)$ \\ \hline
FTN compression factor & $\alpha=0.80$ \\ \hline
Receiver IQI & \begin{tabular}{c} IRR $=17.5$ dB \\ phase mismatch $\phi=10^\circ$ \end{tabular} \\ \hline
Maximum PIC iterations & $I_{\max}=8$ \\ \hline
SD-PIC damping & Adaptive, $\eta_0=0.55$ \\ \hline
SD-PIC parameters & \begin{tabular}{c} $\nu_s=3$, $\eta_{\min}=0.20$ \\ $\gamma_{\rm dec}=0.65$, $\gamma_{\rm inc}=1.05$ \end{tabular} \\ \hline
Variance bounds and cost weight & $\xi_{\min}=10^{-8}$, $\xi_{\max}=2$, $\lambda=0.15$ \\ \hline
PIC stopping threshold & $\varepsilon_{\rm stop}=10^{-5}$ \\ \hline
% \multicolumn{1}{c}{$^\dagger$ Default value unless otherwise specified.} 
\end{tabular}
\end{table}

\section{Simulation Results and Discussion}
\label{sec:simulation}
This section presents Monte Carlo simulation results to evaluate the proposed AFDM-FTN system with receiver IQI and the corresponding WL-MMSE-PIC receivers. The simulations examine four aspects of the proposed framework. First, the overall BER performance is compared among conventional L-MMSE, WL-MMSE, WL-MMSE-HD-PIC, WL-MMSE-SD-PIC, genie aided PIC, and the GAB curve. Second, the iterative behaviors of HD-PIC and SD-PIC are examined to characterize the capability and limitation of PIC based residual interference cancellation. Third, the residual coupling metric and the chirp grid index are investigated to support the local stability and parameter selection analysis. Finally, the impact of the FTN compression factor and the complexity performance tradeoff are discussed to clarify the tradeoff among reliability, efficiency, and complexity for the proposed receivers.

The main simulation parameters are summarized in Table~\ref{tab:sim_params}. Unless otherwise specified, the 4-QAM constellation is employed as the basic constellation with $N=256$. The transmitted symbols are normalized to unit average energy with $\sigma_d^2=1$, and the SNR is defined as $\sigma_d^2/\sigma_w^2$ before receiver IQI. The transmit shaping pulse and receiver matched filtering pulse both adopt root raised-cosine filters with rolloff factor $\beta=0.3$.
% At each SNR point, independent QAM blocks, channel coefficients, Doppler indices, and noise samples are generated, while identical channel and noise realizations are used across all compared receivers for BER evaluation. 
The complex fading coefficients of all paths follow the distribution of $\mathcal{CN}(0,1/P)$. 
The normalized Doppler index for the $p$th path is generated as $\nu_p = \nu_{\max} \cos(\theta)$, where $\nu_{\max}=1$ is the maximum normalized Doppler index and $\theta$ follows a uniform distribution in $[-\pi,\pi]$.
% They are normalized to keep the average channel power unchanged over Monte Carlo trials. Parameters marked by $^\dagger$ denote default settings unless otherwise specified. 
For the receiver IQI model in \eqref{eq:iqi_coeff}, the default setting adopts $\epsilon=0.1$ and $\phi=10^\circ$, which gives $\mathrm{IRR}=10\log_{10}(|\mu|^2/|\nu|^2)\approx17.5$ dB. This setting represents a severe yet practically relevant IQI condition commonly considered in IQI impaired receiver simulations~\cite{tarighat2005iqi}. Similar receiver IQI settings have also been adopted in OFDM direct conversion studies~\cite{tubbax2005iqi}. 

\begin{figure}[!t] 
    \centering
    \includegraphics[width=0.95\columnwidth]{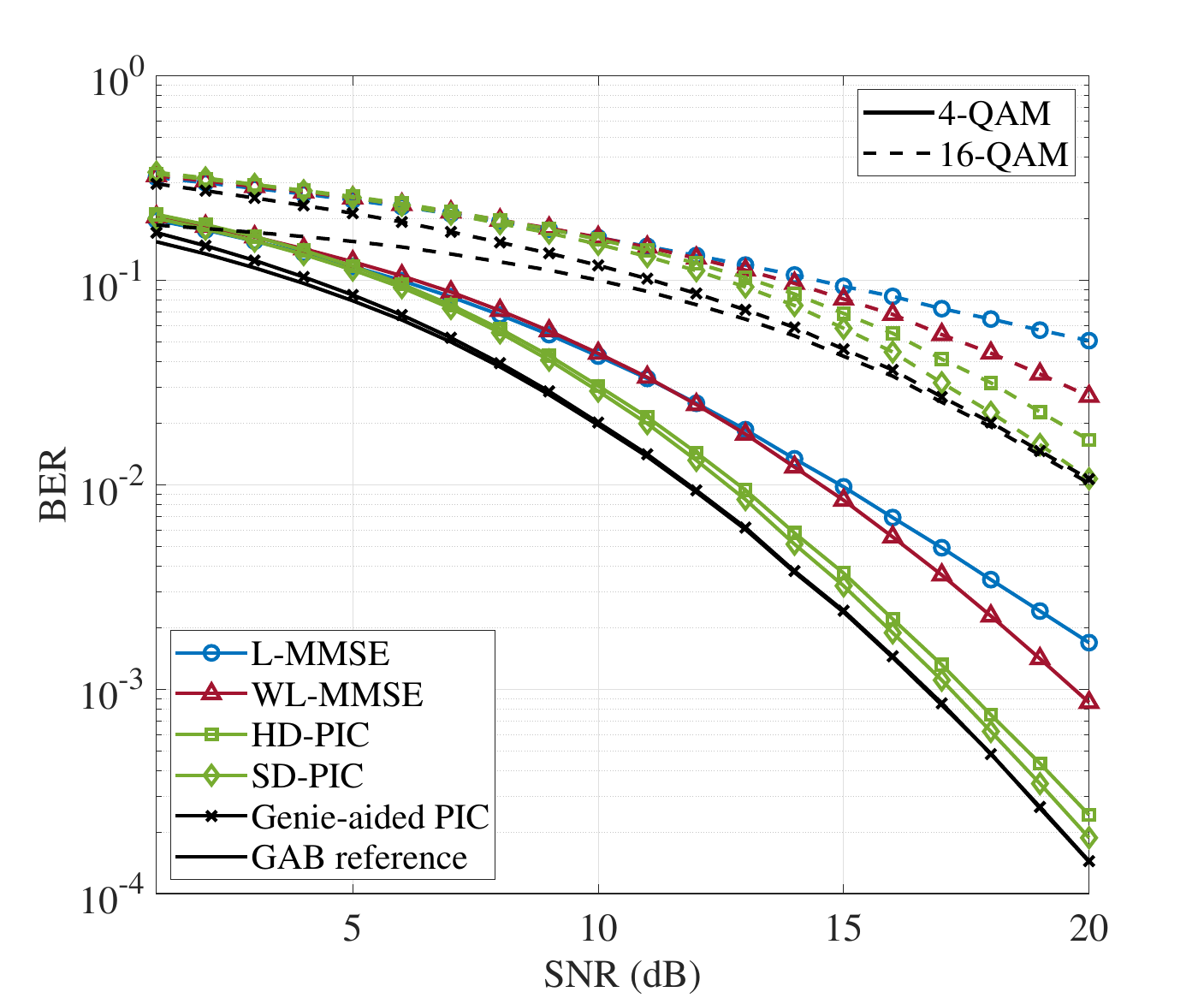}
	\caption{BER comparison of different detectors with 4-QAM and 16-QAM.}
	\label{fig:ber_compare}
\end{figure}

\subsection{BER Performance Comparison}
Fig.~\ref{fig:ber_compare} compares the BER performance of L-MMSE, WL-MMSE, WL-MMSE-HD-PIC, and WL-MMSE-SD-PIC with 4-QAM and 16-QAM under the same channel and receiver IQI settings in Table~\ref{tab:sim_params}, where $\alpha=0.80$, $c_1=3/(2N)$, and receiver IRR $=17.5$ dB. The solid and dashed curves correspond to 4-QAM and 16-QAM, respectively. In addition, two receiver side performance references are included. Genie-aided PIC assumes perfect symbol decisions after WL-MMSE preprocessing and removes the residual coupling using the true transmitted symbols. The GAB curve assumes ideal residual coupling cancellation, and its BER is calculated from the resulting equivalent noise variance under the scalar equivalent SNR approximation.
The conventional L-MMSE detector gives the worst performance because it operates on the nonaugmented model and treats the IQI induced image branch as interference. WL-MMSE exploits the augmented real and image structure of $\mathcal H$, while HD-PIC and SD-PIC further cancel the residual interference associated with the off diagonal post equalization entries. HD-PIC relies on hard residual reconstruction, whereas SD-PIC uses posterior mean feedback together with covariance and pseudo covariance information, resulting in lower BER in the medium- and high SNR regions. For 4-QAM, at the target BER of $2\times10^{-3}$, WL-MMSE, WL-MMSE-HD-PIC, and WL-MMSE-SD-PIC provide SNR gains of about $1.25$ dB, $3.34$ dB, and $3.63$ dB over L-MMSE, respectively. The SD-PIC curve is also within about $0.55$ dB of the GAB reference at the same BER level. The remaining gap is mainly attributed to imperfect feedback, finite iterations, residual uncertainty, and the scalar equivalent SNR approximation used in the analytical reference.
A higher modulation order, i.e., 16-QAM, increases the sensitivity to residual interference and noise, resulting in higher BER than with 4-QAM at the same SNR. Nevertheless, SD-PIC maintains the lowest BER among the practical receivers and approaches the genie aided PIC and GAB reference curves in the high SNR regime. The gain over HD-PIC reflects the benefit of posterior mean feedback and second order reliability information when cancelling residual interference with a higher modulation order.

\begin{figure}[!t]
    \centering
    \includegraphics[width=0.95\columnwidth]{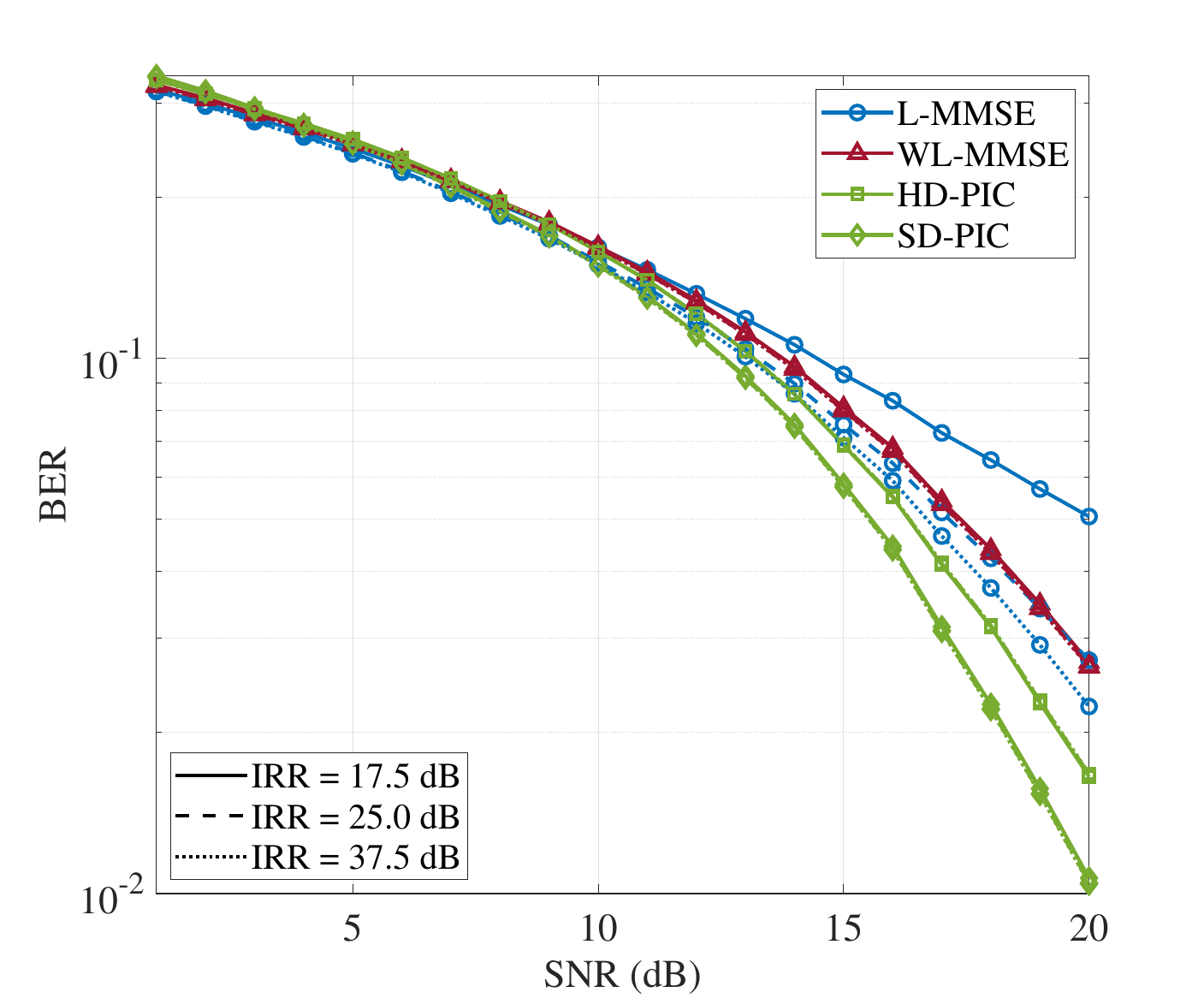}
    \caption{BER comparison of different detectors with 16-QAM under different receiver IRR settings.}
    \label{fig:irr_compare}
\end{figure}
Fig.~\ref{fig:irr_compare} examines the BER performance with 16-QAM under the three receiver IRR settings, while retaining the same channel, FTN compression factor, and detector parameters. For IRR values of $17.5$, $25$, and $37.5$ dB, the progressively weaker image leakage substantially improves the BER of L-MMSE, whose curves exhibit clear separation in the medium and high SNR regions. In contrast, both HD-PIC and SD-PIC exhibit nearly overlapping BER curves across the three settings. Even at IRR $=17.5$ dB, each PIC receiver achieves BER close to its corresponding result at IRR $=37.5$ dB, indicating that augmented preprocessing and iterative residual cancellation effectively suppress the performance loss caused by receiver IQI under the considered conditions. Notably, SD-PIC achieves the lowest BER, followed by HD-PIC and WL-MMSE, in the medium and high SNR regions. The additional SD-PIC gain is consistent with its use of soft feedback and reliability information for residual interference cancellation.

\subsection{Iterative Interference Cancellation Behavior}

\begin{figure*}[!t] 
\centering 
\subfloat[WL-MMSE-HD-PIC.\label{fig:iter_a}]{
\begin{minipage}[t]{0.32\textwidth} 
\centering
\vspace{0pt}
\includegraphics[ 
width=\linewidth
]{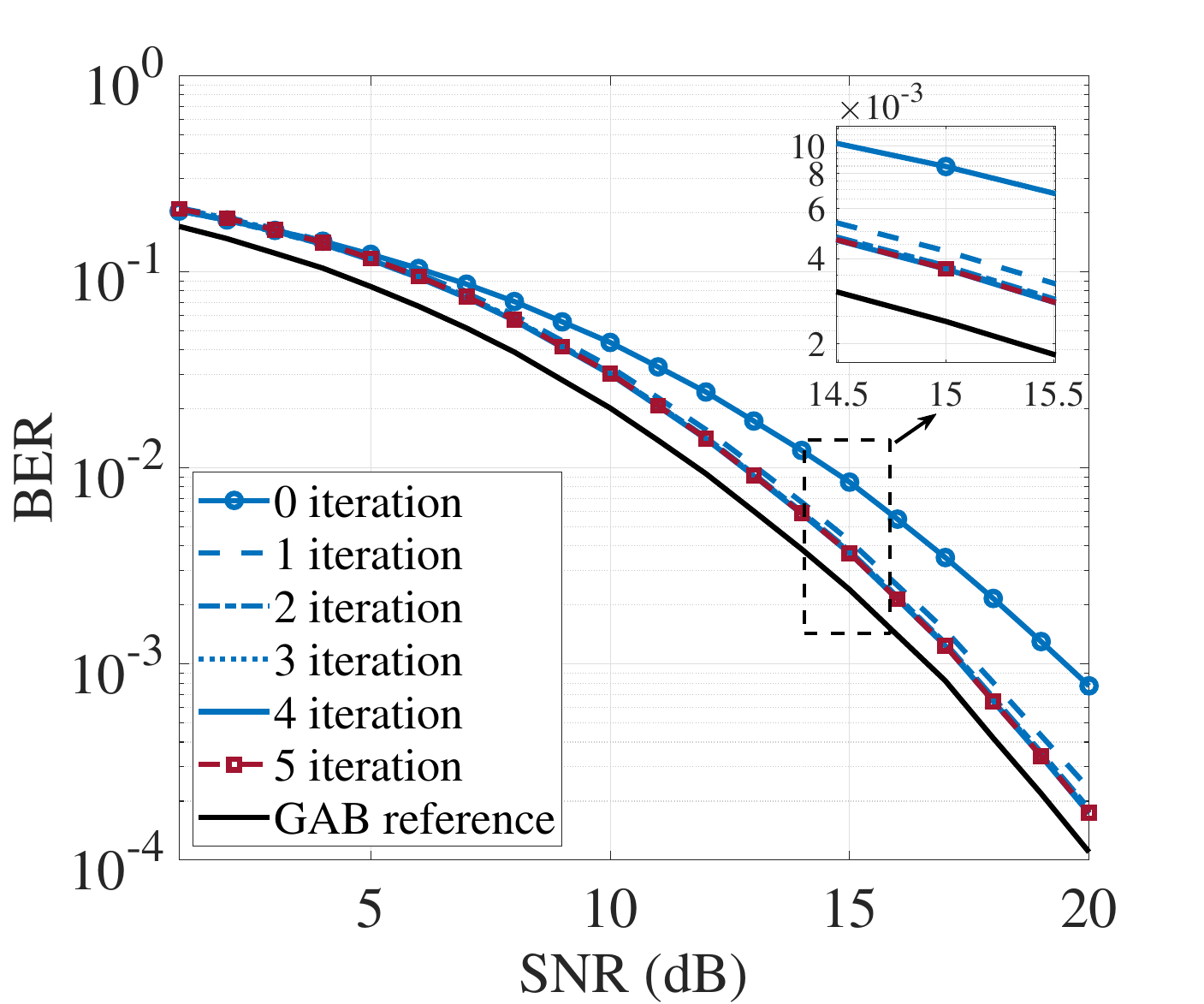}
\end{minipage}}
\hfil
\subfloat[WL-MMSE-SD-PIC.\label{fig:iter_b}]{
\begin{minipage}[t]{0.32\textwidth}
\centering
\vspace{0pt}
\includegraphics[
width=\linewidth
]{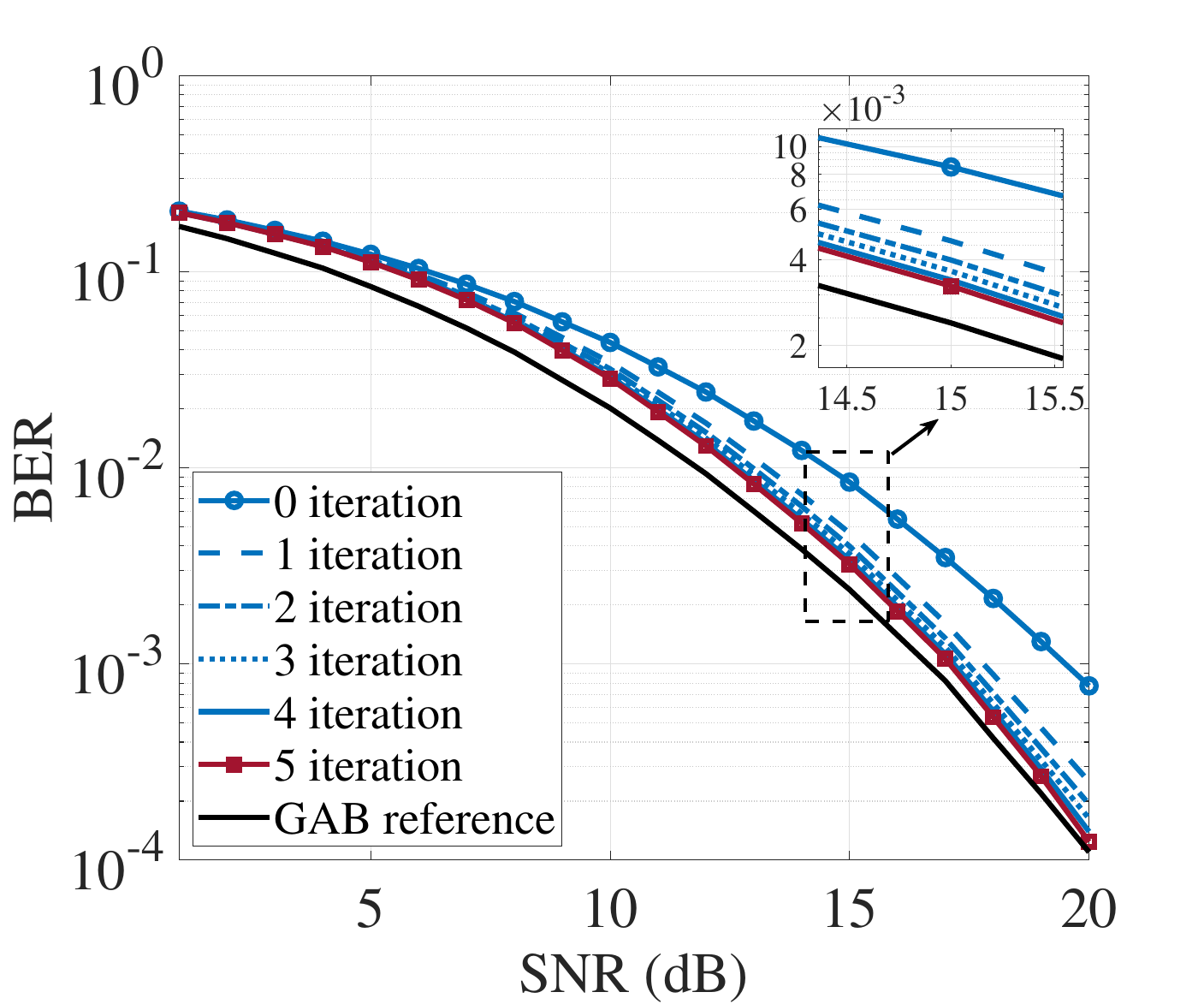}
\end{minipage}}
\hfil
\subfloat[BER versus iteration index at SNR $=16$ dB.\label{fig:iter_c}]{
\begin{minipage}[t]{0.32\textwidth}
\centering
\vspace{0pt}
\includegraphics[ 
width=\linewidth,
]{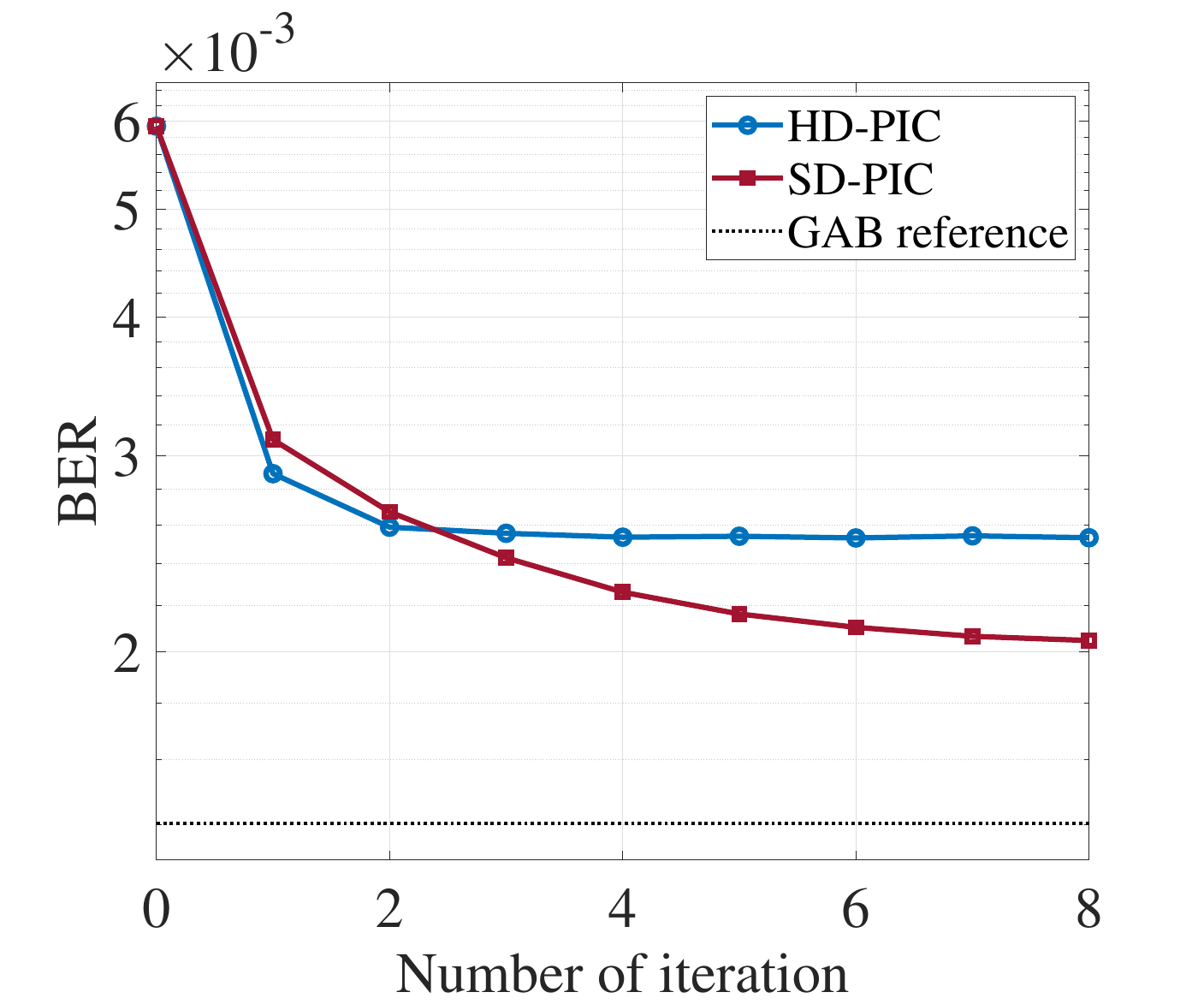}
\end{minipage}}
\caption{Iterative convergence behavior of WL-MMSE-PIC receivers.} \label{fig:iter}
\end{figure*}
Fig.~\ref{fig:iter} evaluates the iterative behavior of the proposed PIC receivers. Fig.~\ref{fig:iter_a} and Fig.~\ref{fig:iter_b} show the BER performance of HD-PIC and SD-PIC when the iteration number is varied from $0$ to $5$, respectively. It can be observed that increasing the iteration number reduces BER for both receivers, especially in the medium to high SNR region. Most of the iteration gain is obtained within the first few iterations, indicating that the dominant residual coupling components are suppressed at the early cancellation stage. After several iterations, the curves tend to saturate, indicating that the remaining performance loss is mainly governed by imperfect feedback and equivalent noise. The iteration gain also depends on the reliability of the feedback symbols. At low SNR, the normalized equivalent noise $\bar{\mathbf v}$ dominates the slicing input in \eqref{eq:hd_norm_recursion}, so tentative decision errors are more likely to be propagated through $\boldsymbol\Gamma_{\rm res}$. Additional iterations then provide limited gain. At high SNR, the feedback becomes more reliable, and SD-PIC approaches the genie aided reference more closely, as shown in Fig.~\ref{fig:iter_b}. This behavior indicates that soft posterior feedback based on covariance and pseudo covariance information reduces the practical loss caused by imperfect residual reconstruction.

Fig.~\ref{fig:iter_c} further compares the BER evolution versus the iteration index at SNR $=16$ dB. HD-PIC converges faster and obtains most of its gain within a small number of iterations, but its steady state BER remains separated from the ideal references because of hard decision error propagation. SD-PIC converges more gradually, while its lower final BER reflects the benefit of reliability aware soft feedback and damping. At SNR $=16$ dB, the BER stabilizes by the seventh iteration. This result illustrates the tradeoff between convergence speed and final detection accuracy.

\begin{figure*}[!t]
\centering
\subfloat[BER versus residual coupling metric under different $\rho_{\rm res}$.\label{fig:residual_coupling_c1_a}]{
\begin{minipage}[t]{0.48\textwidth}
\centering
\includegraphics[width=\linewidth]{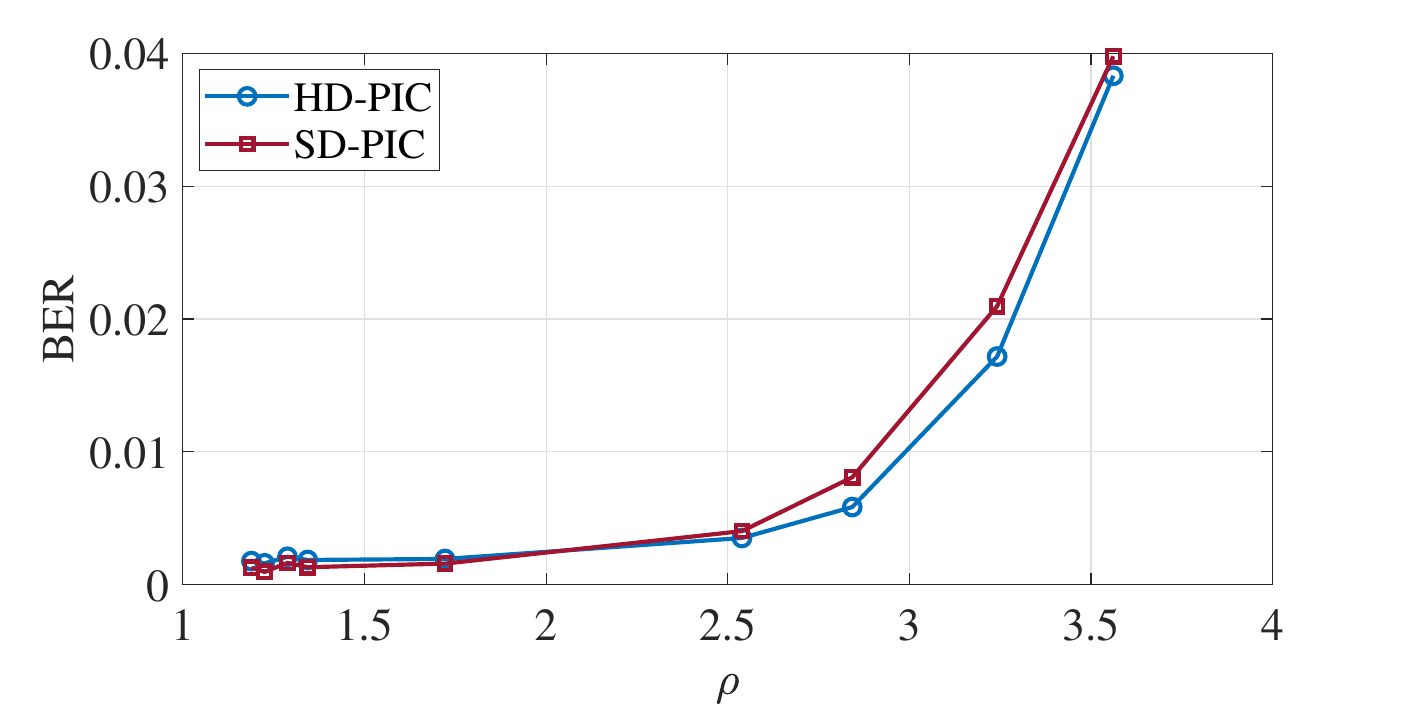}
\end{minipage}}
\hfil
\subfloat[BER versus chirp grid index $q=2Nc_1$.\label{fig:residual_coupling_c1_b}]{
\begin{minipage}[t]{0.48\textwidth}
\centering
\includegraphics[width=\linewidth]{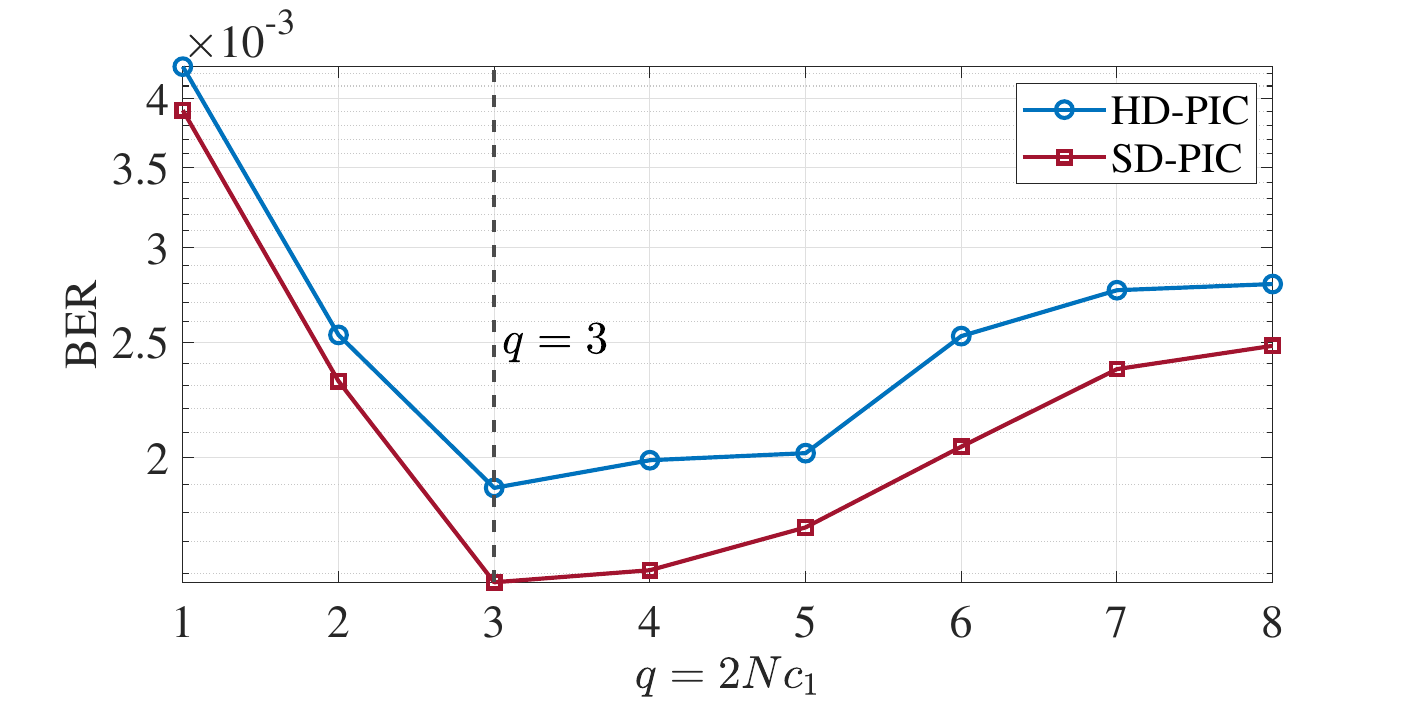}
\end{minipage}}
\caption{Residual coupling validation and chirp parameter selection.}
\label{fig:residual_coupling_c1}
\end{figure*}

Fig.~\ref{fig:residual_coupling_c1} validates the residual coupling interpretation and the chirp parameter selection criterion under the default channel and IQI settings at SNR $=16$ dB. Fig.~\ref{fig:residual_coupling_c1_a} plots BER as a function of the normalized residual coupling metric
$ \rho_{\rm res}=\left\|\boldsymbol\Gamma_{\rm res}\right\|_{\infty}
=\left\|\mathbf D^{-1}\mathbf P_{\rm off}\right\|_{\infty}, $
where $\mathbf D$ and $\mathbf P_{\rm off}$ denote the diagonal and off diagonal components of the post WL-MMSE matrix, respectively. In Fig.~\ref{fig:residual_coupling_c1_a}, $c_1$ is fixed, $\alpha$ is varied from $0.6$ to $1.0$, and the SNR is fixed at $16$ dB. The values below the default $\alpha=0.8$ are included as aggressive FTN compression stress cases. Each operating point is characterized by the corresponding $\rho_{\rm res}$ obtained from the post equalization matrix. A larger $\rho_{\rm res}$ indicates stronger residual coupling and generally leads to higher BER, confirming that $\rho_{\rm res}$ is a useful reliability indicator for PIC based detection. The BER separation between HD-PIC and SD-PIC becomes more evident as $\rho_{\rm res}$ increases, since hard feedback is more susceptible to error propagation whereas soft feedback remains more robust. This trend is consistent with the local stability analysis based on $\boldsymbol\Gamma_{\rm res}$.

Fig.~\ref{fig:residual_coupling_c1_b} shows BER versus the chirp grid index $q=2Nc_1$ at SNR $=16$ dB, where the remaining parameters follow Table~\ref{tab:sim_params}. Under the default simulation setting, the reference profile initialization in Section~\ref{sec:4} gives
$$
q_0=\left\lfloor \frac{N}{D_{\rm eff}(\alpha)+1}\right\rfloor =3.
$$
The BER optimum around $q=3$ is consistent with the residual leakage based grid search, indicating that chirp parameter selection should account for the post equalization coupling structure rather than rely only on waveform parameters. The similar optimal region observed for HD-PIC and SD-PIC further indicates that $c_1$ mainly shapes the residual coupling distribution before the feedback mechanism is applied. Thus, $q=3$ is adopted as the default setting under the considered simulation setup.

\subsection{FTN Compression Factor and Waveform Comparison}
\begin{figure}[!t]
\centering
\includegraphics[
width=0.95\linewidth,
trim=0.2 0.2 0.2 0.1,
clip,
keepaspectratio,
draft = false
]{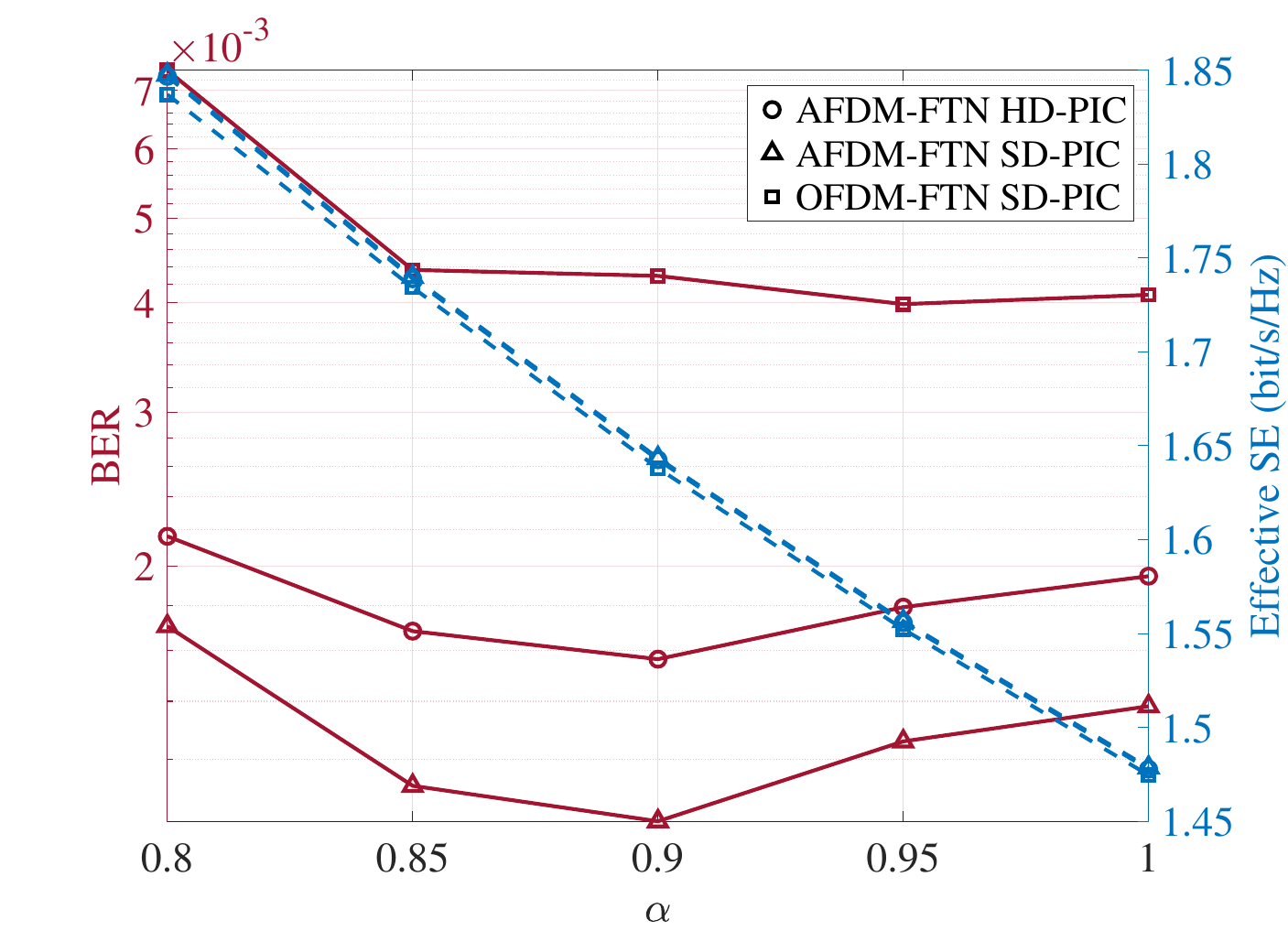}
\caption{BER and effective spectral efficiency under different FTN compression factors.}
\label{fig:ftn_packing_factor}
\end{figure}

Fig.~\ref{fig:ftn_packing_factor} presents the BER performance and effective spectral efficiency under different FTN compression factors at SNR $=16$ dB, where $\alpha$ is swept from $0.80$ to $1.00$ with step size $0.05$. Both AFDM-FTN and OFDM-FTN systems are considered for comparison. The common simulation parameters are $N=256$, $L_{\rm cpp}=10$, $P=3$, $\beta=0.3$, receiver IQI with an IRR of $17.5$ dB and $\phi=10^\circ$, and $I_{\max}=8$. For a fair comparison, both systems use identical transmitted symbols, channel realizations, noise samples, power normalization, FTN pulse, SNR definition, and WL-MMSE-SD-PIC detection. The OFDM-FTN benchmark differs only in the modulation and demodulation transforms, where the DAFT pair is replaced by the conventional IDFT/DFT pair. The left axis reports BER, while the right axis reports the effective spectral efficiency defined by
\begin{equation}
{\eta_{\rm eff}=\frac{N}{N+L_{\rm cpp}}\frac{\log_2 M}{\alpha(1+\beta)}(1-\mathrm{BER}),}
\label{eq:effective_throughput}
\end{equation}
where $N/(N+L_{\rm cpp})$ accounts for CPP overhead, $\log_2M/[\alpha(1+\beta)]$ gives the nominal rate contribution, and $1-\mathrm{BER}$ accounts for detection errors.
The nominal rate contribution increases as $\alpha$ decreases. For the AFDM-FTN SD-PIC curve with 4-QAM and $\beta=0.3$, Fig.~\ref{fig:ftn_packing_factor} shows that the effective spectral efficiency increases from approximately $1.48$ to $1.84$ bit/s/Hz as $\alpha$ decreases from $1.0$ to $0.8$. This rate gain is accompanied by stronger FTN interference, which can degrade BER and reduce effective spectral efficiency.
The BER curves show that this spectral-efficiency gain is accompanied by stronger pulse correlation, which aggravates FTN-ISI and increases the residual coupling after WL-MMSE.
The AFDM-FTN curves achieve lower BER than the OFDM-FTN reference over the considered range of $\alpha$, indicating that DAFT-domain chirp processing is beneficial under the adopted doubly selective channel and IQI settings.
\subsection{Complexity and Performance Tradeoff}
\begin{figure}[!t]
\centering
\includegraphics[
width=0.95\linewidth,
trim=0.2 0.2 0.2 0.1,
clip,
keepaspectratio,
]{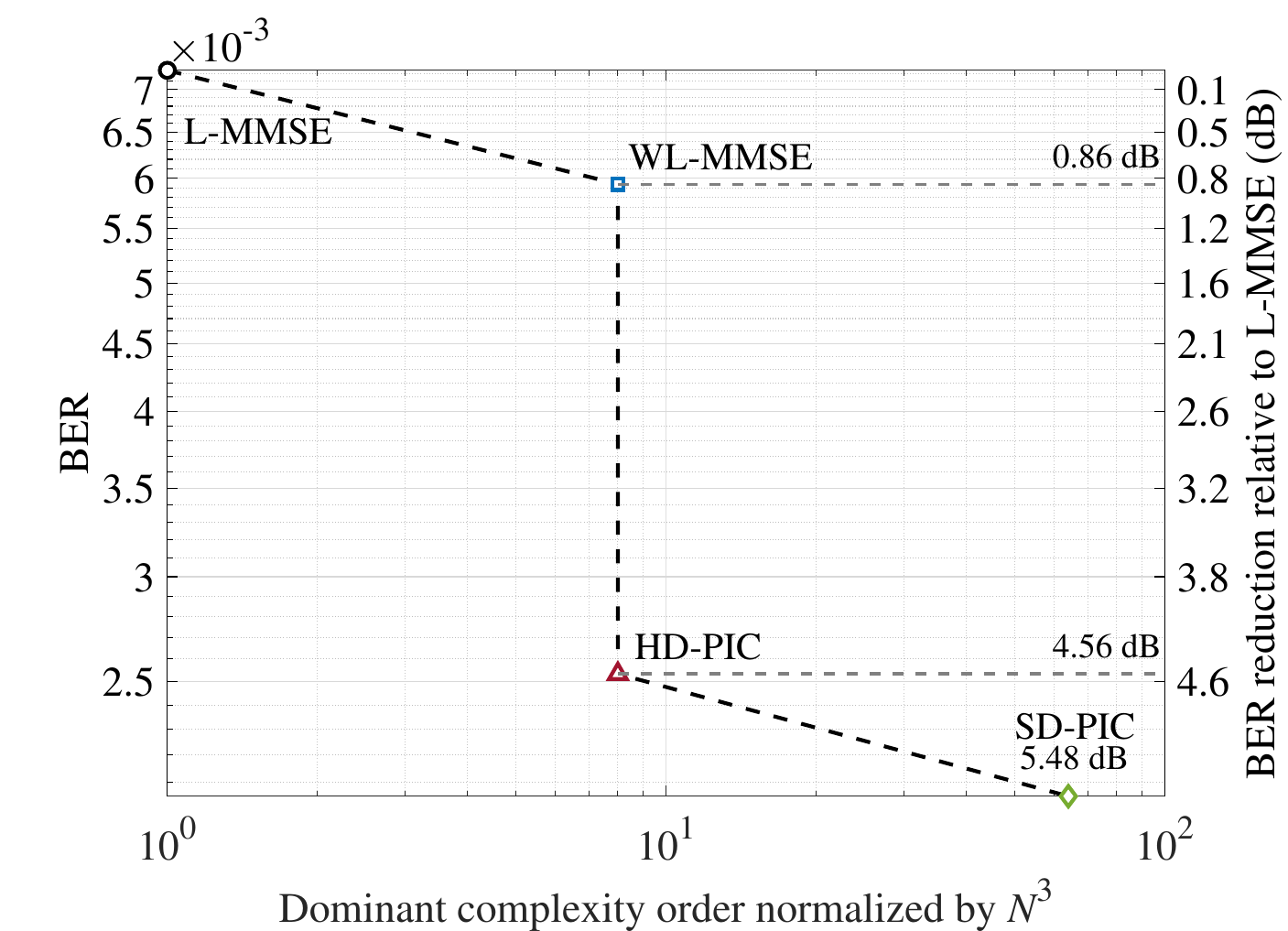}
\caption{Complexity and performance tradeoff at the SNR $=16$ dB.}
\label{fig:complexity_performance}
\end{figure}

Fig.~\ref{fig:complexity_performance} presents the BER performance and computational complexity of the considered receivers at SNR $=16$ dB. The horizontal axis denotes the dominant theoretical complexity normalized by $N^3$, while the right vertical axis gives the BER reduction relative to L-MMSE. Let $P_b$ and $P_b^{\rm L-MMSE}$ denote the BERs of the considered receiver and the L-MMSE receiver, respectively. The BER reduction is defined by
$$
\Delta_{\rm BER}=10\log_{10}\left(\frac{P_b^{\rm L-MMSE}}{P_b}\right).
$$
According to Fig.~\ref{fig:complexity_performance}, WL-MMSE, WL-MMSE-HD-PIC, and WL-MMSE-SD-PIC achieve BER reductions of $0.86$ dB, $4.56$ dB, and $5.48$ dB over L-MMSE, respectively. The WL-MMSE gain comes from augmented real and image processing, while the additional gains of HD-PIC and SD-PIC come from residual post equalization cancellation. Compared with HD-PIC, SD-PIC provides a further $0.92$ dB improvement through posterior mean feedback, covariance and pseudo covariance information, and damping, indicating that reliability information contributes to residual interference reconstruction. The L-MMSE receiver has the lowest complexity but is limited by IQI image leakage and coupled FTN channel interference. WL-MMSE increases the matrix dimension from $N$ to $2N$ and improves BER through augmented processing. HD-PIC adds moderate iterative complexity because its iterations are dominated by residual interference reconstruction and slicing. SD-PIC further adds bivariate posterior probability evaluation, variance and pseudo variance calculation, and second order statistic updates, which increase complexity but give the lowest BER among the practical receivers. Overall, HD-PIC provides a lower complexity operating point, whereas SD-PIC gives a lower BER when the additional computational cost is acceptable.

%%=======================================Chapter 6 Conclusion ============================================%%
\section{Conclusion}
\label{sec:conclusion}

This paper studied receiver design for AFDM-FTN systems under doubly selective propagation and receiver IQI. A unified augmented input/output model was first established to jointly characterize FTN-ISI, doubly selective channel spreading, and image leakage induced by IQI. Based on this model, a WL-MMSE-PIC receiver was developed, where WL-MMSE preprocessing suppresses the real and image components in the augmented domain, while HD-PIC and SD-PIC further cancel the residual post equalization coupling. Theoretical analysis related the iterative behavior to the normalized residual coupling matrix and to the covariance and pseudo covariance blocks of the SD-PIC feedback error, derived a genie aided reference under ideal residual coupling cancellation, and discussed the influence of the AFDM chirp parameter $c_1$ on the interference distribution after equalization. Simulation results demonstrated that the proposed WL-MMSE-HD-PIC and WL-MMSE-SD-PIC receivers achieve lower BER than conventional L-MMSE and augmented WL-MMSE detectors, and that receiver oriented chirp parameter selection further improves BER performance. These results demonstrate that the proposed WL-MMSE-PIC framework provides an analytically tractable receiver solution for IQI impaired AFDM-FTN systems.

%%=======================================References ============================================%%

\end{document}